\documentclass{article}
\usepackage[T1]{fontenc}
\usepackage[letterpaper, margin=2cm]{geometry}
\usepackage{amsmath, amsthm, amssymb}
\usepackage{mathtools}
\usepackage{epigraph}
\usepackage{graphicx}
\usepackage[inline]{enumitem}
\usepackage[noend]{algpseudocode}
\usepackage{float} 
\usepackage[dvipsnames]{xcolor} 
\usepackage{tikz}
\usepackage[framemethod=tikz]{mdframed}
\usepackage{environ, varwidth}
\usepackage{multicol}
\usepackage{hyperref}
\usepackage{bm}
\usepackage{trimclip}
\usepackage{newtxtext, newtxmath}
\usepackage{pifont} 
\usepackage{subdepth} 

\hypersetup{
    colorlinks,
    linkcolor={blue},
    citecolor={blue},
    urlcolor={blue}
}

\newtheoremstyle{customstyle}
  {}
  {}
  {}
  {}
  {\bfseries}
  {.}
  { }
  {\thmname{#1}\thmnumber{ #2}\thmnote{ (#3)}}%
\theoremstyle{customstyle}

\usepackage{xpatch}
\makeatletter
\xpatchcmd{\endmdframed}
  {\aftergroup\endmdf@trivlist\color@endgroup}
  {\endmdf@trivlist\color@endgroup\@doendpe}
  {}{}
\makeatother

\newmdtheoremenv[innertopmargin=-2pt, skipabove=5pt, skipbelow=0pt, linewidth=2pt, linecolor=Green, backgroundcolor=Green!10, bottomline=false, leftline=false, rightline=false]{thesis}{Thesis}
\newmdtheoremenv[innertopmargin=-2pt, skipabove=5pt, skipbelow=0pt, linewidth=2pt, linecolor=Green, backgroundcolor=Green!10, bottomline=false, leftline=false, rightline=false]{theorem}{Theorem}
\newmdtheoremenv[innertopmargin=-2pt, skipabove=5pt, skipbelow=0pt, linewidth=2pt, linecolor=Green, backgroundcolor=Green!10, bottomline=false, leftline=false, rightline=false]{lemma}{Lemma}
\newmdtheoremenv[innertopmargin=-2pt, skipabove=5pt, skipbelow=0pt, linewidth=2pt, linecolor=Green, backgroundcolor=Green!10, bottomline=false, leftline=false, rightline=false]{corollary}{Corollary}
\newmdtheoremenv[innertopmargin=-2pt, skipabove=5pt, skipbelow=0pt, linewidth=2pt, linecolor=Yellow, backgroundcolor=Yellow!10, bottomline=false, leftline=false, rightline=false]{definition}{Definition}

\newenvironment{prf}{\begin{mdframed}[skipabove=5pt, backgroundcolor=Gray!10, topline=false, bottomline=false, leftline=false, rightline=false]\begin{proof}}{\end{proof}\end{mdframed}}
\newenvironment{prf-sketch}{\begin{mdframed}[skipabove=5pt, backgroundcolor=Gray!10, topline=false, bottomline=false, leftline=false, rightline=false]\begin{proof}}{\end{proof}\end{mdframed}}

\newenvironment{algo}{\begin{samepage}\medskip\hrule\begin{algorithmic}}{\end{algorithmic}\hrule\medskip\end{samepage}}

\newcommand{\qu}[1]{\mathsf{#1}}
\newcommand{\uq}[1]{\text{$#1$}}

\newcommand{\fm}[1]{\mathsf{#1}}
\newcommand{\fmc}[1]{\mathtt{#1}}

\newcommand{\textscsf}[1]{\textsf{\textsc{#1}}}
\DeclareMathOperator{\Con}{\textscsf{Con}}
\DeclareMathOperator{\Pvb}{\textscsf{Pvb}}
\DeclareMathOperator{\PvbG}{\Pvb_\fm{G}}
\DeclareMathOperator{\Chk}{\textscsf{Ck}}

\DeclareMathOperator{\pc}{\mathit{pv}}

\DeclareMathOperator{\nil}{\mathit{nil}}
\DeclareMathOperator{\inc}{\mathit{inc}}
\DeclareMathOperator{\prj}{\mathit{prj}}
\DeclareMathOperator{\eq}{\mathit{eq}}
\DeclareMathOperator{\add}{\mathit{add}}
\DeclareMathOperator{\mul}{\mathit{mul}}
\DeclareMathOperator*{\cat}{\mathrel{\scalebox{0.6}[0.7]{$\frown$}}}
\DeclareMathOperator*{\Cat}{\mathrel{\scalebox{0.6}[0.7]{$\raisebox{0.25ex}{$\frown$}\mkern-17mu\raisebox{-0.25ex}{$\frown$}$}}}
\DeclareMathOperator*{\argmin}{arg\,min}
\DeclareMathOperator{\fcheck}{\mathit{check}}
\DeclareMathOperator{\fsim}{\mathit{sim}}

\newcommand{\rep}[1]{\text{\guilsinglleft$#1$\guilsinglright}}

\newcommand{\fmtick}{\emph{\textquotesingle}}
\newcommand{\fmeq}{\mathrel{\resizebox{1.15\height}{1.15\height}{=}}}
\newcommand{\T}{\textbf{\textsf{T}}}
\newcommand{\Q}{\textbf{\textsf{Q}}}
\newcommand{\PA}{\textbf{\textsf{PA}}}
\newcommand{\A}{\textbf{\textsf{A}}}
\newcommand{\B}{\textbf{\textsf{B}}}
\newcommand{\cmark}{\text{\small \ding{51}}}
\newcommand{\xmark}{\text{\small \ding{55}}}
\newcommand{\imp}{\mathrel{\scalebox{0.8}[0.8]{\clipbox{1.75mm 0mm 0mm 0mm}{$\rightarrow$}}}}

\newbox\numBoxA
\newdimen\numCornerHgt
\setbox\numBoxA=\hbox{\footnotesize $\ulcorner$}
\global\numCornerHgt=\ht\numBoxA
\newdimen\numArgHgt
\def\num #1{%
\setbox\numBoxA=\hbox{$#1$}%
\numArgHgt=\ht\numBoxA%
\ifnum     \numArgHgt<\numCornerHgt \numArgHgt=0pt%
\else \advance \numArgHgt by -\numCornerHgt%
\fi \raise\numArgHgt\hbox{\footnotesize $\ulcorner$} \box\numBoxA %
\raise\numArgHgt\hbox{\footnotesize $\urcorner$}}

\newcommand{\numrep}[1]{\num{\hspace{-3.33pt}\rep{#1}\hspace{-3.33pt}}}

\begin{document}

\title{\vspace{-1.5cm}What If Turing Had Preceded Gödel?}
\author{$\fm{S}$ebastian $\fm{0}$berhoff\\{\small oberhoff.sebastian@gmail.com}}
\date{\today}

\maketitle

\begin{abstract}
  The overarching theme of the following pages is that mathematical logic---centered around the incompleteness theorems---is first and foremost an investigation of \emph{computation}, not arithmetic. Guided by this intuition we will show the following.
  \begin{itemize}
    \item First, we'll all but eliminate the need for Gödel numbers.
    \item Next, we'll introduce a novel notational device for representable functions and walk through a condensed demonstration that Peano Arithmetic can represent every computable function. It has achieved Turing completeness.
    \item Continuing, we'll derive the Diagonal Lemma and First Incompleteness Theorem using significantly simplified proofs.
    \item Approaching the Second Incompleteness Theorem, we'll be able to use some self-referential trickery to avoid much of the technical morass surrounding it; arriving at three separate versions.
    \item Extending the analogy between the First Incompleteness Theorem and the Unsolvability of the Halting Problem produces an equivalent of the Nondeterministic Time Hierarchy Theorem from the field of computational complexity.
    \item Lastly, we'll briefly peer into the realm of the uncomputable by connecting our ideas to oracles.
  \end{itemize}
\end{abstract}

\epigraph{In March of 1977, I met the great AI pioneer Marvin Minsky for the first time. It was an unforgettable experience. One of the most memorable remarks he made to me was this one: ``Gödel should just have thought up Lisp; it would have made the proof of his theorem much easier.'' I knew exactly what Minksy meant by that, I could see a grain of truth in it, and moreover I knew it had been made with tongue semi in cheek.}{\textit{Douglas Hofstadter}}

\begin{center}
  \includegraphics[scale=0.115]{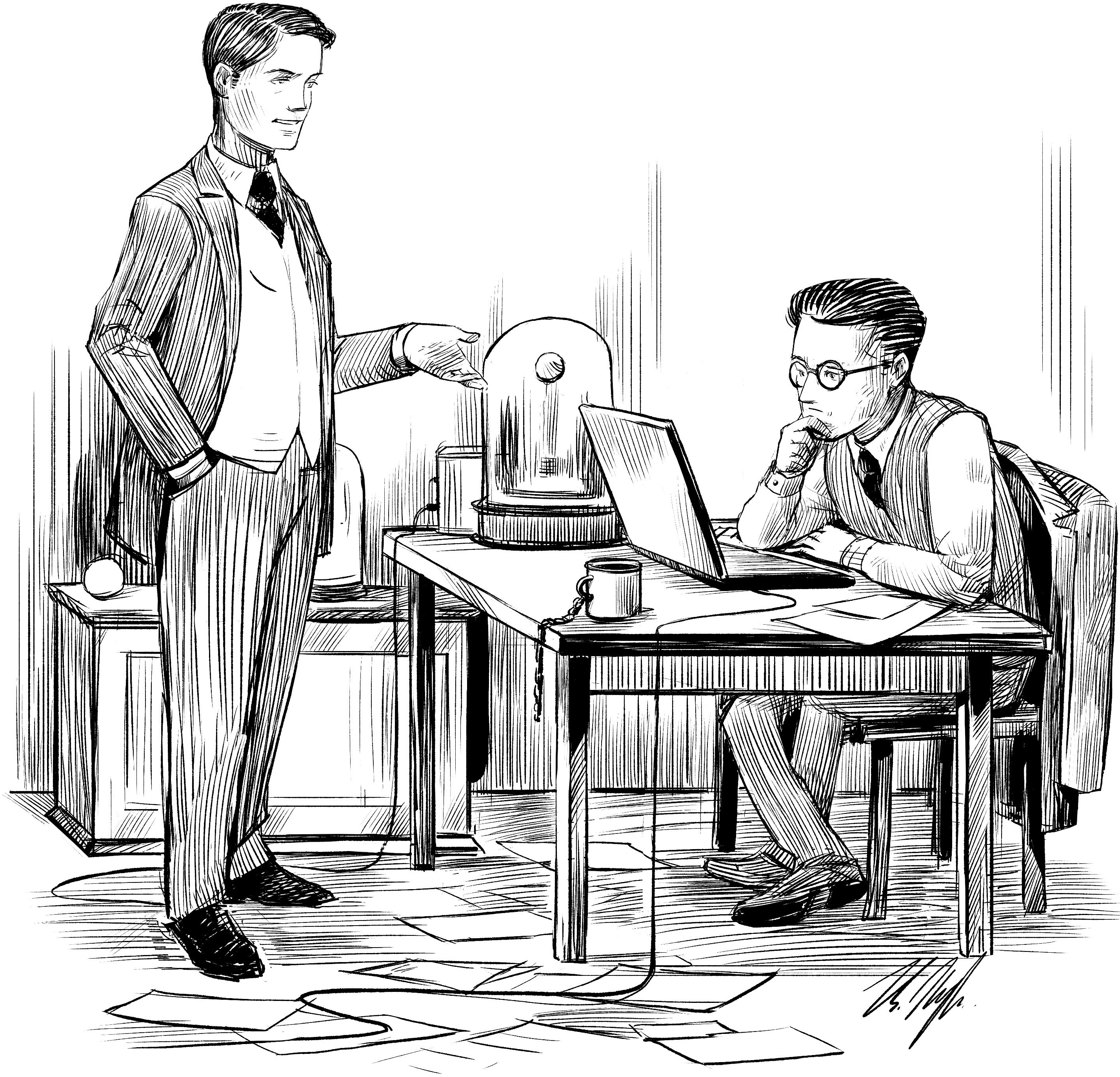}
  \newline
  \footnotesize{Illustration by ManBroDude}
\end{center}

\section{Introduction}

Ever since 1931 the main protagonist in any telling of mathematical logic has always been Kurt Gödel and his First Incompleteness Theorem. Roughly speaking this theorem states that any sensible, sufficiently powerful formal system is incomplete; it'll contain a statement that is neither provable nor disprovable.

However, while the statement of the theorem is quite simple, the proof is a very different matter. Entire books, both technical and non-technical, have been devoted to it over the years; some of which, I'm assuming in the sequel, you're already familiar with. Where does this complexity come from? And what can be done about it? These are questions I already addressed in \cite{oberhoff}, though in a rather inexact manner. Indeed, back then I explicitly left important details to ``smarter people'' in the conclusion. Now, some time later, and hopefully a little smarter, I want to return to the question and provide a more precise answer myself.

To that end let's start with a high-level overview of the famous 1931 paper titled ``On Formally Undecidable Propositions Of Principia Mathematica And Related Systems'' \cite{goedel}. In this paper Gödel dealt with a formal system he had adapted from work by Bertrand Russel and Alfred North Whitehead published in three volumes of the \emph{Principia Mathematica} between 1910 and 1913. This system was based on type theory, a rather tedious framework which has since fallen out of favor. But its central pillar---Peano Arithmetic---has stood the test of time. It defines the natural numbers and is built as follows.

\begin{definition}[Theory]
  A theory is a set of formal sentences which is closed under logical consequence. Furthermore, if $\T\fmtick$ and $\T$ are theories such that $\T\fmtick \subseteq \T$, then we'll say that $\T$ is an \emph{extension} of $\T\fmtick$. And if $\fm{A}$ is a sentence such that $\fm{A} \in \T$, then we'll instead write $\T \vdash \fm{A}$ which reads as ``$\T$ proves $\fm{A}$''.

  We'll only consider first-order theories with equality.
\end{definition}

For practical purposes ``being a logical consequence'' is going to mean ``derivable using informal natural deduction'' to us.

\begin{definition}[Robinson Arithmetic]
  The theory of \emph{Robinson Arithmetic}, or $\Q$ for short, is defined from the following axioms.
  \begin{description}
    \item[Axioms Of Successorship]
          \hfill\newline
          \begin{minipage}{0.225\textwidth}
            \noindent\smallskip
            \[
              \forall \fm{x} \colon \neg\, \fm{S}\fm{x} \fmeq \fm{0} \tag{S0}\label{S0}
            \]
          \end{minipage}
          \quad
          \begin{minipage}{0.3\textwidth}
            \noindent\smallskip
            \[
              \forall \fm{x}, \fm{y} \colon \fm{S}\fm{x} \fmeq \fm{S}\fm{y} \rightarrow \fm{x} \fmeq \fm{y} \tag{S1}\label{S1}
            \]
          \end{minipage}
          \quad
          \begin{minipage}{0.3\textwidth}
            \noindent\smallskip
            \[
              \forall \fm{x} \colon \fm{x} \fmeq \fm{0} \lor \exists \fm{y} \colon \fm{S}\fm{y} \fmeq \fm{x} \tag{S2}\label{S2}
            \]
          \end{minipage}
    \item[Axioms Of Arithmetic]
          \hfill\newline
          \begin{minipage}{0.4\textwidth}
            \noindent
            \begin{align*}
              \forall \fm{x} \colon         & \fm{x} + \fm{0} \fmeq \fm{x} \tag{A0}\label{A0}                        \\
              \forall \fm{x}, \fm{y} \colon & \fm{x} + \fm{S}\fm{y} \fmeq \fm{S}(\fm{x} + \fm{y}) \tag{A1}\label{A1}
            \end{align*}
          \end{minipage}
          \begin{minipage}{0.4\textwidth}
            \noindent
            \begin{align*}
              \forall \fm{x} \colon         & \fm{x} \times \fm{0} \fmeq \fm{0} \tag{M0}\label{M0}                                \\
              \forall \fm{x}, \fm{y} \colon & \fm{x} \times \fm{S}\fm{y} \fmeq \fm{x} + (\fm{x} \times \fm{y}) \tag{M1}\label{M1}
            \end{align*}
          \end{minipage}
  \end{description}
\end{definition}

\begin{definition}[Peano Arithmetic]
  The theory of \emph{Peano Arithmetic}, or $\PA$ for short, is defined using all the axioms of $\Q$ as well as the following axiom schema.
  \begin{description}
    \item[Induction]
          $(\fm{P}(\fm{0}) \land \forall \fm{x} \colon \fm{P}(\fm{x}) \rightarrow \fm{P}(\fm{S}\fm{x})) \rightarrow \forall \fm{x} \colon \fm{P}(\fm{x})$ for any formula $\fm{P}$.
  \end{description}
\end{definition}
The main reason $\Q$ is interesting is that it is a finitely axiomatized and fairly weak theory. It can't even prove the commutativity of addition. Yet it still suffices to prove many of our results. So separating it out from $\PA$ allows us to state our theorems in their sharpest form.

With all that in mind, here are the major steps in Gödel's 1931 proof:
\begin{description}
  \item[Gödel Numbering]
        Akin to modern-day ASCII encoding, Gödel first turned every string and list of strings into a number.
  \item[Computability]
        Using the model of computation known as primitive recursive functions Gödel then implemented a proof verifier in full detail, showing that $\PA$ is computably enumerable.\footnote{I am in full agreement with Robert I. Soare \cite{soare} that ``computably enumerable'' should replace ``recursively enumerable''.}
  \item[Representation]
        Next, Gödel briefly pointed out that $\PA$ could represent---that is reason about---primitive recursive functions, and thereby its own proof verifier.
  \item[Diagonalization]
        Finally, having equipped $\PA$ with the ability of introspection, Gödel diagonalized against provability-in-$\PA$. This produced a sentence asserting its own unprovability, leading to the First Incompleteness Theorem: $\PA$ can't decide every sentence.
  \item[Re-Representation]
        Almost as an afterthought, Gödel noted that the proof of the First Incompleteness Theorem could be carried out within $\PA$. By a simple argument this gave his Second Incompleteness Theorem: $\PA$ couldn't prove its own consistency.
\end{description}
This, in essence, is still the argument given to this day. And we're not going to modify its basic structure either. Rather we'll take these items one by one and see where simplifications can be made.

\section{Gödel Numbering}

While today there is an unending variety of computational models, stretching from Diophantine equations to the RISC-V instruction set architecture, back in 1931 there were only very few available options. That's why when Gödel drew the primitive recursive functions from the shelf the very first challenge he faced was a simple type error. The formal system he was working with was built out of strings over an exotic character set whereas primitive recursive functions only mapped natural numbers to natural numbers. If one wanted to devise a proof verifier in this model of computation, some type conversion would be unavoidable. So that's exactly what Gödel did using a scheme that since has become known as Gödel numbering.

Yet driven primarily by necessity as opposed to elegance, Gödel ended up constructing a rather roundabout numbering involving products of prime numbers raised to various powers. Considering the time at which Gödel was writing this is obviously quite understandable. What's more difficult to explain is the enduring popularity of Gödel's original scheme. After all, there's a much more natural mapping: read every formal string as a natural number that has been written in base $k$ where $k$ is the size of our alphabet. To see this idea in action just consider hexadecimal notation where the letters ``A''--``F'' are used to represent the numbers 10--15. Furthermore, sequences of strings can be handled much more straightforwardly by introducing a new-line character.

Looking at it from this angle there's really no Gödel numbering left worth mentioning. It's just a change of script. And if prime numbers ever turn out to be more convenient later on, then all that's required is to add a pre-processor which converts from the powers-of-$k$ to the powers-of-primes encoding. Doing so can be done even using primitive recursive functions without encountering major obstacles.

Okay, so maybe that is a little simpler, you might say. But \emph{strictly speaking} isn't this still a Gödel numbering? Sure. That is unless you take it one step further. One only has to reinterpret the formal symbol ``$\fm{0}$'' to indicate the first string and ``$\fm{S}$'' to indicate the lexicographical successor function. At that point Gödel numbers are truly gone.

However, as tempting as this approach towards unification may be, it also harbors the potential for a lot of confusion. Setting up string addition and string multiplication etc., while technically unobjectionable, brings with it significant mental overhead, simply because we're not used to treating strings in this fashion. So instead of treating everything as a string let's go the other way and treat everything as a number. In particular, we'll regard our formal strings as already being natural numbers. These numbers are written using the conventional symbols of the language of arithmetic. And the corner quotes usually employed to denote Gödel numbers will be used to point directly to numerals instead.

\subsection{Numerals}

\begin{definition}
  When handling collections of items we'll employ boldface variables to avoid clutter. E.g., $\bm{x} \coloneqq x_1, \dots, x_k$.
\end{definition}

\begin{definition}[Numeral]
  Let $x \in \mathbb{N}$. Then
  \[
    \num{x} \enskip\coloneqq\enskip \qu{S \dots S0}
  \]
  where $\qu{S}$ is repeated $x$ times. We'll call $\num{x}$ the \emph{numeral} of $x$. Also, $\num{\bm{x}}$ is to be taken element-wise.
\end{definition}

\begin{definition}[Ordering]
  The $<$ relation is defined in terms of addition as follows.
  \begin{align*}
    \fm{x} < \fm{y}    & \enskip\coloneqq\enskip \exists \fm{z} \colon \fm{S} \fm{z} + \fm{x} \fmeq \fm{y}, \\
    \fm{x} \leq \fm{y} & \enskip\coloneqq\enskip \fm{x} < \fm{y} \lor \fm{x} \fmeq \fm{y}.
  \end{align*}
\end{definition}

Now is a good time to collect our first batch of basic lemmas.

\begin{lemma}[Monotonicity Of Successorship]\label{lm-mono}
  \[
    \Q \vdash \forall \fm{x}, \fm{y} \colon \fm{x} < \fm{y} \leftrightarrow \fm{S}\fm{x} < \fm{S}\fm{y}.
  \]
\end{lemma}

\begin{prf}
  Let $\fmc{x}, \fmc{y}$ and $\fmc{z}$ be arbitrary. Then we can see
  \[
    \fm{S}\fmc{z} + \fmc{x} \fmeq \fmc{y} \,\leftrightarrow\, \fm{S}(\fm{S}\fmc{z} + \fmc{x}) \fmeq \fm{S}\fmc{y} \quad\text{and}\quad \fm{S}(\fm{S}\fmc{z} + \fmc{x}) \fmeq \fm{S}\fmc{y} \,\leftrightarrow\, \fm{S}\fmc{z} + \fm{S}\fmc{x} \fmeq \fm{S}\fmc{y}
  \]
  by Leibniz's law (substitution of equals), \ref{S1}, and \ref{A1}. This of course contracts to $\fmc{x} < \fmc{y} \leftrightarrow \fm{S}\fmc{x} < \fm{S}\fmc{y}$.
\end{prf}

\begin{lemma}[Totality Of Ordering With Respect To Numerals]\label{lm-total-num}
  For any $n \in \mathbb{N}$:
  \[
    \Q \vdash \forall \fm{x} \colon \fm{x} < \num{n} \lor \num{n} < \fm{x} \lor \fm{x} \fmeq \num{n}.
  \]
\end{lemma}

\begin{prf}
  By induction on $n$.
  \begin{description}
    \item[Base Case]
          Suppose $\num{n} = \fm{0}$. And let $\fmc{x}$ be arbitrary. Then by \ref{S2} we have two cases to consider.
          \begin{description}
            \item[$\fmc{x} \fmeq \fm{0}$]
                  Then trivially $\fmc{x} \fmeq \num{n}$.
            \item[$\fmc{x} \fmeq \fm{S}\fmc{y}$]
                  Here $\fm{S}\fmc{y} + \num{n} \fmeq \fm{S}\fmc{y} + \fm{0} \fmeq \fm{S}\fmc{y} \fmeq \fmc{x}$ by \ref{A0} and hence $\num{n} < \fmc{x}$.
          \end{description}
    \item[Induction]
          Suppose $\num{n} = \fm{S}\num{m}$ for some $m \in \mathbb{N}$. And let $\fmc{x}$ be arbitrary again. Then once more by \ref{S2}.
          \begin{description}
            \item[$\fmc{x} \fmeq \fm{0}$]
                  Then $\fm{S}\num{m} + \fmc{x} \fmeq \fm{S}\num{m} + \fm{0} \fmeq \fm{S}\num{m} \fmeq \num{n}$ by \ref{A0} and hence $\fmc{x} < \num{n}$.
            \item[$\fmc{x} \fmeq \fm{S}\fmc{y}$]
                  Here we have $\fmc{y} < \num{m} \lor \num{m} < \fmc{y} \lor \fmc{y} \fmeq \num{m}$ by inductive hypothesis, whereby $\fmc{x} < \num{n} \lor \num{n} < \fmc{x} \lor \fmc{x} \fmeq \num{n}$ using \hyperref[lm-mono]{monotonicity} and Leibniz's law.\qedhere
          \end{description}
  \end{description}
\end{prf}

\begin{lemma}[Non-Negativity Lemma]\label{lm-noneg}
  \[
    \Q \vdash \forall \fm{x} \colon \neg\, \fm{x} < \fm{0}.
  \]
\end{lemma}

\begin{prf}
  By way of contradiction, let $\fmc{x}$ be an arbitrary object such that $\fmc{x} < \fm{0}$. Then by definition we can obtain a $\fmc{z}$ such that $\fm{S}\fmc{z} + \fmc{x} \fmeq \fm{0}$. We now use \ref{S2} to distinguish two cases.
  \begin{description}
    \item[$\fmc{x} \fmeq \fm{0}$]
          Then by \ref{A0} we have $\fm{S}\fmc{z} \fmeq \fm{S}\fmc{z} + \fm{0} \fmeq \fm{S}\fmc{z} + \fmc{x} \fmeq \fm{0}$ which contradicts \ref{S0}.
    \item[$\fmc{x} \fmeq \fm{S}\fmc{y}$]
          Then by \ref{A1} we have $\fm{S}(\fm{S}\fmc{z} + \fmc{y}) \fmeq \fm{S}\fmc{z} + \fm{S}\fmc{y} \fmeq \fm{S}\fmc{z} + \fmc{x} \fmeq \fm{0}$ which again contradicts \ref{S0}.\qedhere
  \end{description}
\end{prf}

\begin{definition}[Bounded Quantifier]
  We'll employ the following shorthand for any formula $\fm{P}$:
  \begin{align*}
    \forall \fm{x} < \fm{y} \colon \fm{P}(\fm{x}, \fm{y}) & \quad\coloneqq\quad \forall \fm{x} \colon \fm{x} < \fm{y} \rightarrow \fm{P}(\fm{x}, \fm{y}), \\
    \exists \fm{x} < \fm{y} \colon \fm{P}(\fm{x}, \fm{y}) & \quad\coloneqq\quad \exists \fm{x} \colon \fm{x} < \fm{y} \land \fm{P}(\fm{x}, \fm{y}).
  \end{align*}
  The same goes for $\leq$. Such a quantifier is also referred to as a \emph{bounded quantifier}.
\end{definition}

\begin{lemma}[Bounded Quantification Lemma]\label{lm-bounded}
  For any $n \in \mathbb{N}$:
  \[
    \Q \vdash \forall \fm{x} < \fm{S}\num{n} \colon \fm{x} \fmeq \fm{0} \lor \dots \lor \fm{x} \fmeq \num{n}.
  \]
\end{lemma}

\begin{prf}
  By induction on $n$.
  \begin{description}
    \item[Base Case]
          Suppose $\num{n} = \fm{0}$. Then we need to establish $\fmc{x} < \fm{S}\fm{0} \rightarrow \fmc{x} \fmeq \fm{0}$ for an arbitrary $\fmc{x}$. By \ref{S2} there are two cases.
          \begin{description}
            \item[$\fmc{x} \fmeq \fm{0}$]
                  This case is trivial.
            \item[$\fmc{x} \fmeq \fm{S}\fmc{y}$]
                  This assumption allows us to rule out the antecedent. By way of contradiction, assume $\fmc{x} < \fm{S}\fm{0}$. Then $\fmc{y} < \fm{0}$ by \hyperref[lm-mono]{monotonicity}. But this contradicts \hyperref[lm-noneg]{non-negativity}.
          \end{description}
    \item[Induction]
          Suppose $\num{n} = \fm{S}\num{m}$ for some $m \in \mathbb{N}$. And let $\fmc{x}$ be arbitrary again. Then one last time by \ref{S2}.
          \begin{description}
            \item[$\fmc{x} \fmeq \fm{0}$]
                  This case is trivial.
            \item[$\fmc{x} \fmeq \fm{S}\fmc{y}$]
                  Then $\fmc{x} < \fm{S}\num{n}$ implies $\fmc{y} < \fm{S}\num{m}$ by \hyperref[lm-mono]{monotonicity}. Therefore, by the inductive hypothesis, $\fmc{y} \fmeq \fm{0} \lor \dots \lor \fmc{y} \fmeq \num{m}$. And so $\fmc{x} \fmeq \fm{S}\fm{0} \lor \dots \lor \fmc{x} \fmeq \num{n}$.\qedhere
          \end{description}
  \end{description}
\end{prf}

\section{Computability}

This section is the most infamous in Gödel's original proof. It's a chain of 46 definitions, up to 3 lines each, spread over four pages, altogether implementing a proof verifier for $\PA$ using primitive recursive functions.\footnote{Gödel actually gave a relation here, not a function. But that muddles the boundary between the different sections, so I prefer to stick to functions here.} Really though, it's nothing more than an painstaking demonstration that proofs for $\PA$ could be checked by a computer. Back in 1931 mathematicians may have had every right to be doubtful of this claim. This was long before we built actual computers and familiarized ourselves with high-level programming languages. And most importantly, it was before the field of computer science fully dawned in 1936 with the universal acceptance of the following thesis.

\begin{thesis}[Church-Turing Thesis]
  Every computable function can be computed by a Turing machine (or equivalently by a $\lambda$-calculus expression, or by a general recursive function, etc.).
\end{thesis}

The exact status of the Church-Turing Thesis is still debated. Is it a definition? Is it a law of nature? Regardless, for us it's going to be nothing more than a bridge between intuition and rigor. Rather than having to spill pages upon pages of ink writing code in the assembly language of general recursive functions, we're going to allow ourselves to write simple pseudocode. At that point we'll exclaim ``clearly, that's computable'', and then continue on under the assumption that, if called upon, we could always compile our pseudocode down into any desired universal model of computation.

Using this modern perspective we can then distill this section into its essence like so.

\begin{definition}[$\xmark$, $\cmark$]
  The symbols ``$\xmark$'' and ``$\cmark$'' indicate rejection and acceptance respectively. They can be taken to be arbitrary constants (e.g. 0 and 1). Their only important feature is that they're distinct.
\end{definition}

\begin{definition}[Proof Verifier]
  A function $\pc \colon \mathbb{N}^2 \to \{\xmark, \cmark\}$ is called a \emph{proof verifier} for some theory $\T$ if whenever $\T \vdash \fm{A}$, then $\pc(p, \fm{A}) = \cmark$ for some $p \in \mathbb{N}$, and otherwise $\pc(p, \fm{A}) = \xmark$.
\end{definition}

\begin{definition}[Computably Enumerable Theory]
  A theory $\T$ is considered to be \textit{computably enumerable} if there exists a computable proof verifier for $\T$.
\end{definition}

\begin{theorem}
  $\Q$ and $\PA$ are computably enumerable theories.
\end{theorem}

\begin{prf-sketch}
The following implementation of $\pc$ will do.
\begin{algo}
  \Function{$\pc(p, x)$}{}
  \For{$l \in \text{ lines of } p$}
  \If{$l$ is not an axiom \textbf{and} $l$ doesn't follow from the previous lines}
  \State \Return $\xmark$
  \EndIf
  \EndFor
  \If{the last line of $p$ isn't $x$}
  \State \Return $\xmark$
  \EndIf
  \State \Return $\cmark$
  \EndFunction
\end{algo}
\end{prf-sketch}

\section{Representation}

Next, we turn to an extremely important section of the proof; the representation of computable functions as formulas. Roughly speaking, for a function $f$ to be represented in a theory $\T$ means that if $f(x) = y$, then $\T$ can prove this fact. And it does so by use of a single formula called ``the representation of $f$''.

\begin{definition}[Unique Existence]
  For any formula $\qu{P}$
  \[
    \exists! \fm{x} \colon \fm{P}(\fm{x}) \quad\coloneqq\quad \exists \fm{x} \colon \fm{P}(\fm{x}) \land \forall \fm{y} \colon \fm{P}(\fm{y}) \rightarrow \fm{y} \fmeq \fm{x}.
  \]
\end{definition}

\begin{definition}[Representable Function]
  Let $\T$ be a theory containing the axioms of successorship. Then a partial function $f \colon \mathbb{N}^k \to \mathbb{N}$ is said to be \emph{represented} in $\T$ if there exists a formula $\rep{f}$ with $k+1$ free variables such that whenever $f(\bm{x}) = y$ then
  \[
    \T \vdash \exists! \fm{y} \colon \rep{f}(\num{\bm{x}}, \fm{y}) \land \rep{f}(\num{\bm{x}}, \num{y}).
  \]
\end{definition}
Informally the above sentence says ``$f(\bm{x})$ \emph{only} outputs $y$''.

As already stated, the concept of representability will be absolutely central for the remainder of this presentation. But notationally it's still a bit of a mouthful. We're frequently going to find ourselves wanting to assert that the output of $f(\bm{x})$ obeys some predicate $\fm{P}$. Wouldn't it be nice if we could make this assertion by simply writing $\fm{P}(f(\num{\bm{x}}))$? We may not be able to write $f$ directly into our formulas. But no matter, we can simply declare $\fm{P}(f(\num{\bm{x}}))$ to be shorthand for $\exists \fm{y} \colon \rep{f}(\num{\bm{x}}, \fm{y}) \land \fm{P}(\fm{y})$.

This is something other people have already come up with and they've christened such appearances of $f(\num{\bm{x}})$ ``pseudo-terms''. However, while it's a step in the right direction, pseudo-terms still have some significant shortcomings. Most importantly, because pseudo-terms don't guarantee the uniqueness of $f(\num{\bm{x}})$'s output Leibniz's law breaks down. This means that if one derives $f(\num{\bm{x}}) \fmeq g(\num{\bm{z}})$, then one can't easily conclude $\fm{P}(f(\num{\bm{x}})) \leftrightarrow \fm{P}(g(\num{\bm{z}}))$. Most grievously, this has the consequence that equalities of pseudo-terms don't enjoy transitivity.

The situation brings to mind a wise quote by Alfred North Whitehead: ``By relieving the brain of all unnecessary work, a good notation sets it free to concentrate on more advanced problems, and in effect increases the mental power of the race.'' In this vein, what follows may appear as a bit of a diversion. But it'll pay untold dividends in the long run.

\begin{definition}[Representable Function Symbol \& Representable Term]
  Let $f$ be represented in $\T$. Then we'll allow ``$f$'' to appear in our formal strings anywhere a regular function symbol could appear. We'll refer to such function symbols as \emph{representable function symbols}. And terms containing representable function symbols are called \emph{representable terms}.

  In order to disabbreviate representable function symbols from some expression we proceed as follows. Let our expression be of the form $\fm{P}(f(\bm{\fm{r}}))$ where $\fm{P}$ is the smallest sub-formula containing $f$ and $f$ is one of the innermost occurrences of a representable function symbol, meaning $\bm{\fm{r}}$ is a collection of regular terms. Then we perform the following expansion:
  \[
    \fm{P}(f(\bm{\fm{r}})) \quad\coloneqq\quad (\exists! \fm{y} \colon f(\bm{\fm{r}}, \fm{y})) \land (\exists \fm{y} \colon f(\bm{\fm{r}}, \fm{y}) \land \fm{P}(\fm{y})).
  \]
  And then we continue expanding $\fm{P}(\fm{y})$ in this way until all representable function symbols have been eliminated.

  Also, it is important that representable terms are expanded \emph{before} newly introduced predicate symbols are.
\end{definition}

That last stipulation is sadly necessary in order to resolve the following ambiguity. (This issue is already present with pseudo-terms.) Suppose we've defined $\fm{P}(\fm{y}) \coloneqq \neg\, \fm{y} \fmeq \num{\cmark}$ and $f$ is some representable function. Then naively $\fm{P}(f(\num{\bm{x}}))$ might have two possible expansions:
\[
  \underset{\text{(correct)}}{(\exists! \fm{y} \colon \rep{f}(\num{\bm{x}}, \fm{y})) \land (\exists \fm{y} \colon \rep{f}(\num{\bm{x}}, \fm{y}) \land \neg \fm{y} \fmeq \num{\cmark})} \quad\text{or}\quad \underset{\text{(incorrect)}}{\neg ((\exists! \fm{y} \colon \rep{f}(\num{\bm{x}}, \fm{y})) \land (\exists \fm{y} \colon \rep{f}(\num{\bm{x}}, \fm{y}) \land  \fm{y} \fmeq \num{\cmark}))}.
\]
The left side says that $f(\bm{x})$ outputs something other than $\cmark$. The right side says that $f(\bm{x})$ doesn't output $\cmark$. But that might be because $f(\bm{x})$ doesn't output anything at all. Some choice had to be made. And the former struck me as more intuitive and useful. This tripwire is also the reason I'm avoiding the use of $\neq$ in formulas.

\begin{lemma}
  Representable terms expand equivalently in any order.
\end{lemma}

\begin{prf}
  Let's consider the case where we're dealing with just two competing representable function symbols: $\fm{P}(f(\bm{\fm{r}}), g(\bm{\fm{s}}))$ with $\fm{P}$ being some formula and $\bm{\fm{r}}$ and $\bm{\fm{s}}$ being collections of regular terms. Then the expansions we need to compare are
  \[
    (\exists! \fm{y} \colon f(\bm{\fm{r}}, \fm{y})) \land (\exists \fm{y} \colon f(\bm{\fm{r}}, \fm{y}) \land ((\exists! \fm{z} \colon g(\bm{\fm{s}}, \fm{z})) \land (\exists \fm{z} \colon g(\bm{\fm{s}}, \fm{z}) \land \fm{P}(\fm{y}, \fm{z}))))\phantom{.}
  \]
  and
  \[
    (\exists! \fm{z} \colon g(\bm{\fm{r}}, \fm{z})) \land (\exists \fm{z} \colon g(\bm{\fm{r}}, \fm{z}) \land ((\exists! \fm{y} \colon f(\bm{\fm{s}}, \fm{y})) \land (\exists \fm{y} \colon f(\bm{\fm{s}}, \fm{y}) \land \fm{P}(\fm{y}, \fm{z})))).
  \]
  These can be checked to be equivalent by going through all four combinations of setting $\exists! \fm{y} \colon f(\bm{\fm{r}}, \fm{y})$ and $\exists! \fm{z} \colon g(\bm{\fm{s}}, \fm{z})$ to true or false.

  The general case for an arbitrary number of representable terms can then be obtained by observing that any permutation can be decomposed into a series of transpositions.
\end{prf}

\begin{lemma}[Leibniz's Law For Representable Terms]\label{lm-leibniz}
  Let $\fm{r}$ and $\fm{s}$ be closed representable terms. And let $\fm{P}$ be a formula with one free variable. Then the following is logically valid:
  \[
    \fm{r} \fmeq \fm{s} \rightarrow (\fm{P}(\fm{r}) \leftrightarrow \fm{P}(\fm{s})).
  \]
\end{lemma}

\begin{prf}
  By induction on the number of representable function symbols appearing in $\qu{r} \fmeq \qu{s}$.
  \begin{description}
    \item[Base Case]
          If $\fm{r}$ and $\fm{s}$ are regular terms, then the statement is trivial.
    \item[Induction]
          So let's assume that the statement holds for up to $n$ representable functions and that $\fm{r} \fmeq \fm{s}$ contains $n + 1$ representable function symbols. Furthermore, without loss of generality let's focus only on the case where $\fm{r}$ contains a representable function symbol, meaning $\fm{r}$ is of the form $\fm{r}\fmtick(f(\bm{\fm{t}}))$ for some open representable term $\fm{r}\fmtick$ and some collection of closed regular terms $\bm{\fm{t}}$. Then the desired statement expands to
          \[
            (\exists! \fm{y} \colon \rep{f}(\bm{\fm{t}}, \fm{y})) \land (\exists \fm{y} \colon \rep{f}(\bm{\fm{t}}, \fm{y}) \land \fm{r}\fmtick(\fm{y}) \fmeq \fm{s}) \rightarrow (\exists! \fm{y} \colon \rep{f}(\bm{\fm{t}}, \fm{y})) \land (\exists \fm{y} \colon \rep{f}(\bm{\fm{t}}, \fm{y}) \land (\fm{P}(\fm{r}\fmtick(\fm{y})) \leftrightarrow \fm{P}(\fm{s}))).
          \]
          This can be simplified into
          \[
            \fm{r}\fmtick(\fmc{y}) \fmeq \fm{s} \rightarrow (\fm{P}(\fm{r}\fmtick(\fmc{y})) \leftrightarrow \fm{P}(\fm{s}))
          \]
          where $\fmc{y}$ is the unique object such that $\rep{f}(\bm{\fm{t}}, \fmc{y})$. At this stage, $\fm{r}\fmtick(\fmc{y}) = \fm{s}$ is closed and contains only $n$ representable function symbols. So we can bring the inductive hypothesis to bear which concludes the argument.\qedhere
  \end{description}
\end{prf}

\begin{lemma}[Rules Of Equality For Representable Terms]\label{lm-eq}
  Let $\fm{r}, \fm{s}$, and $\fm{t}$ be closed representable terms. Then the following rules of equality for representable terms are logically valid.
  \begin{description}[leftmargin=!, labelwidth=\widthof{Transitivity:}]
    \item[Symmetry] $\fm{r} \fmeq \fm{s} \rightarrow \fm{s} \fmeq \fm{r}$.
    \item[Transitivity] $\fm{r} \fmeq \fm{s} \rightarrow (\fm{s} \fmeq \fm{t} \rightarrow \fm{r} \fmeq \fm{t})$.
  \end{description}
\end{lemma}

\begin{prf}
  \phantom{}
  \begin{description}
    \item[Symmetry]
          Here we can expand notation, use ordinary symmetry, and then contract notation again.
    \item[Transitivity]
          By \hyperref[lm-leibniz]{Leibniz's Law} we have $\fm{r} \fmeq \fm{s} \rightarrow (\fm{s} \fmeq \fm{t} \leftrightarrow \fm{r} \fmeq \fm{t})$ which implies the desired statement.\qedhere
  \end{description}
\end{prf}

Reflexivity is conspicuously absent in the these rules. This could be remedied by replacing the central ``$\land$'' in the definition of representable terms with a ``$\rightarrow$''. But that brings its own issues. For example, one can then no longer use representable terms to introduce existential quantifiers. Besides, we're going to be able to introduce equalities using the following lemma instead.

\begin{lemma}[Runtime Lemma]\label{lm-run}
  Let $f$ be a function which is represented in $\T$. And suppose $f(\bm{x}) = y \neq z$. Then
  \[
    \T \vdash f(\num{\bm{x}}) \fmeq \num{y} \quad\text{and}\quad \T \vdash \neg\, f(\num{\bm{x}}) \fmeq \num{z}.
  \]
\end{lemma}

\begin{prf}
  After expanding notation and minor simplification $f(\num{\bm{x}}) \fmeq \num{y}$ can be turned into
  \[
    \exists ! \fm{y} \colon \rep{f}(\num{\bm{x}}, \fm{y}) \land \rep{f}(\num{\bm{x}}, \num{y}),
  \]
  which is provable in $\T$ by the definition of representable functions. This also allows $\T$ to disprove $f(\num{\bm{x}}) \fmeq \num{z}$ since if it were true, then by transitivity we'd have $\num{y} \fmeq \num{z}$. And that is false by the axioms of successorship alone.
\end{prf}

\subsection{Turing Completeness of $\Q$}

There are of course many kinds of functions out there which one might want to represent. But for us there's one kind in particular which is of primary concern: the computable functions. A theory which represents all computable functions has in a quite literal sense achieved Turing completeness. It has a fully-fledged programming language embedded within it. And a computable function's representation is none other than its source code in this language. This is why I chose the notation $\rep{f}$, as it's reminiscent of how the code for a Turing machine $M$ is often denoted $\langle M \rangle$. Furthermore, in order to ``execute'' a representation $\rep{f}$ on input $\bm{x}$ one can, for example, go through all possible strings and check if any of them proves a theorem of the form $f(\num{\bm{x}}) \fmeq \num{y}$ for some $y$. If so, $y$ is the output. Naturally, one would hope that this isn't the most efficient way of running a representation. But it proves the point.

This allows us to make the central observation of this section: $\Q$ (and consequently $\PA$) represents every computable function. The way I like to think of this is that the axioms of successorship describe what data is and the axioms of arithmetic delineate a universal gate set. To check this claim we're going to need a Turing complete model of computation. The following one is particularly convenient.

\begin{definition}[Arithmetic Functions]
  The following functions of type $\mathbb{N}^k \to \mathbb{N}$ are arithmetic functions.
  \begin{itemize}
    \item $\nil(x) = 0$
    \item $\inc(x) = x + 1$
    \item $\prj_{k,i}(x_1, \dots, x_k) = x_i$
    \item $\eq(x, y) = \begin{cases} 1 & \text{if } x = y,\\ 0 & \text{otherwise}\end{cases}$
    \item $\add(x, y) = x + y$
    \item $\mul(x, y) = x \times y$
  \end{itemize}
  Furthermore, if $g, g_1, \dots, g_l$ are arithmetic functions, then the following are also arithmetic functions.
  \begin{itemize}
    \item $f(\bm{x}) = g_0(g_1(\bm{x}), \dots, g_l(\bm{x}))$
    \item $f(\bm{x}) = \argmin_y g(\bm{x}, y) = 0$
  \end{itemize}
\end{definition}

This model of computation is almost identical to the general recursive functions. The only difference is that we've replaced recursion with addition and multiplication. As Gödel pointed out towards the end of his 1931 paper, this model of computation is in fact able to emulate recursion. That allows it to inherit computational universality from the general recursive functions. But it is much easier to represent.

\begin{theorem}\label{thm-qrep}
  $\Q$ represents every computable function.
\end{theorem}

\begin{prf}
  We begin by using the Church-Turing Thesis to assert that any computable function can be constructed using arithmetic functions. This allows us to proceed by structural induction, stating the representation to be used and then verifying it in turn.
  \begin{description}
    \item[Base Cases]
          \phantom{}
          \begin{description}
            \item[Nil]
                  \[
                    \rep{\nil}(x, y) \enskip\coloneqq\enskip \fm{0} \fmeq \fm{y}
                  \]
                  Suppose $\nil(x) = y$. Then $\num{y} = \num{\xmark} = \fm{0}$. Thus, trivially $\Q \vdash \exists! \fm{y} \colon \fm{0} \fmeq \fm{y} \land \fm{0} \fmeq \fm{0}$.
            \item[Increment]
                  \[
                    \rep{\inc}(\fm{x}, \fm{y}) \enskip\coloneqq\enskip \fm{S}\fm{x} \fmeq \fm{y}
                  \]
                  Suppose $\inc(x) = y$. Then $\num{y} = \fm{S}\num{x}$. Thus, trivially $\Q \vdash \exists! \fm{y} \colon \fm{S}\num{x} \fmeq \fm{y} \land \fm{S}\num{x} \fmeq \fm{S}\num{x}$.
            \item[Projection]
                  \[
                    \rep{\prj_{k,i}}(\fm{x}_1, \dots, \fm{x}_k, \fm{y}) \enskip\coloneqq\enskip \fm{x}_i \fmeq \fm{y}
                  \]
                  Suppose $\prj_{k,i}(\bm{x}) = y$. Then $\num{y} = \num{x_i}$. Thus, trivially $\Q \vdash \exists! \fm{y} \colon \num{x_i} \fmeq \fm{y} \land \num{x_i} \fmeq \num{x_i}$.
            \item[Equality]
                  \[
                    \rep{\eq}(\fm{x}, \fm{y}, \fm{z}) \enskip\coloneqq\enskip (\fm{x} \fmeq \fm{y} \land \fm{z} \fmeq \fm{S}\fm{0}) \lor (\neg\, \fm{x} \fmeq \fm{y} \land \fm{z} \fmeq \fm{0})
                  \]
                  Suppose $\eq(x, y) = z$. Then we have two cases.
                  \begin{description}
                    \item[$\num{x} = \num{y}$ and $\num{z} = \fm{S}\fm{0}$]
                          Here the sought claim reduces to the once again trivial
                          \[
                            \Q \vdash \exists! \fm{z} \colon \num{x} \fmeq \num{y} \land \fm{z} \fmeq \fm{S}\fm{0} \land \num{x} \fmeq \num{y} \land \fm{S}\fm{0} \fmeq \fm{S}\fm{0}.
                          \]
                    \item[$\num{x} \neq \num{y}$ and $\num{z} = \fm{0}$]
                          Here the sought claim reduces to the once again trivial
                          \[
                            \Q \vdash \exists! \fm{z} \colon \neg\, \num{x} \fmeq \num{y} \land \fm{z} \fmeq \fm{0} \land \neg\, \num{x} \fmeq \num{y} \land \fm{0} \fmeq \fm{0}.
                          \]
                  \end{description}
            \item[Addition]
                  \[
                    \rep{\add}(\fm{x}, \fm{y}, \fm{z}) \enskip\coloneqq\enskip \fm{x} + \fm{y} \fmeq \fm{z}
                  \]
                  We first show by induction on $y$ that $\Q \vdash \num{x} + \num{y} \fmeq \num{x + y}$ for any $x, y \in \mathbb{N}$.
                  \begin{description}
                    \item[Base Case]
                          $\Q \vdash \num{x} + \fm{0} \fmeq \num{x}$ follows immediately from \ref{A0}.
                    \item[Induction]
                          Let $y = z + 1$ so that $\num{y} = \fm{S}\num{z}$. Then
                          \[
                            \Q \vdash \num{x} + \num{y} \fmeq \fm{S}(\num{x} + \num{z}) \fmeq \fm{S}\num{x + z} \fmeq \num{x + y}
                          \]
                          by \ref{A1}, followed by the inductive hypothesis, and finally simple equality.
                  \end{description}
                  Suppose $\add(x, y) = z$. Then $\num{z} = \num{x + y}$. Thus, after substituting $\num{x} + \num{y} \fmeq \num{x + y}$, we trivially get
                  \[
                    \Q \vdash \exists! \fm{z} \colon \num{x} + \num{y} \fmeq \fm{z} \land \num{x} + \num{y} = \num{x + y}.
                  \]
            \item[Multiplication]
                  \[
                    \rep{\mul}(\fm{x}, \fm{y}, \fm{z}) \enskip\coloneqq\enskip \fm{x} \times \fm{y} \fmeq \fm{z}
                  \]
                  Similar to addition, we first show by induction on $y$ that $\Q \vdash \qu{\num{x} \times \num{y} \fmeq \num{x \times y}}$ for any $x, y \in \mathbb{N}$.
                  \begin{description}
                    \item[Base Case]
                          $\Q \vdash \num{x} \times \fm{0} \fmeq \fm{0}$ follows immediately from \ref{M0}.
                    \item[Induction]
                          Let $y = z + 1$ so that $\num{y} = \fm{S}\num{z}$. Then
                          \[
                            \Q \vdash \num{x} \times \num{y} \fmeq \num{x} + (\num{x} \times \num{z}) \fmeq \num{x} + \num{x \times z} \fmeq \num{x + x \times z} \fmeq \num{x \times y}
                          \]
                          by \ref{M1}, followed by the inductive hypothesis, the previously established fact that $\Q \vdash \num{x} + \num{y} \fmeq \num{x + y}$, and finally simple equality.
                  \end{description}
                  Suppose $\mul(x, y) = z$. Then $\num{z} = \num{x \times y}$. Thus, after substituting $\num{x} \times \num{y} \fmeq \num{x \times y}$, we trivially get
                  \[
                    \Q \vdash \exists! \fm{z} \colon \num{x} \times \num{y} \fmeq \fm{z} \land \num{x} \times \num{y} = \num{x \times y}.
                  \]
          \end{description}
    \item[Inductions]
          \phantom{}
          \begin{description}
            \item[Composition]
                  Here we use the inductive hypothesis to assume that $g_0, \dots, g_l$ are already represented which allows us to represent $f$ as follows.
                  \[
                    \rep{f}(\bm{\fm{x}}, \fm{y}) \enskip\coloneqq\enskip g_0(g_1(\bm{\fm{x}}), \dots, g_l(\bm{\fm{x}})) \fmeq \fm{y}
                  \]
                  Suppose $f(\bm{x}) = y$. Then for some $y_1, \dots, y_l$ we must have
                  \[
                    g_1(\bm{x}) = y_1,\quad \dots,\quad g_l(\bm{x}) = y_l, \quad\text{and}\quad g_0(y_1, \dots, y_l) = y.
                  \]
                  So by the \hyperref[lm-run]{Runtime Lemma}
                  \[
                    \Q \vdash g_1(\num{\bm{x}}) \fmeq \num{y_1},\quad \dots,\quad \Q \vdash g_l(\num{\bm{x}}) \fmeq \num{y_l}, \quad\text{and}\quad \Q \vdash g_0(\num{y_1}, \dots, \num{y_l}) \fmeq \num{y}.
                  \]
                  Thus, we can take the trivial $\Q \vdash \qu{\exists! y \colon \num{y} \fmeq y \land \num{y} \fmeq \num{y}}$ and repeatedly replace representable terms using \hyperref[lm-leibniz]{Leibniz's Law} until we arrive at
                  \[
                    \Q \vdash \exists! \fm{y} \colon g_0(g_1(\num{\bm{x}}), \dots, g_l(\num{\bm{x}})) \fmeq y \land g_0(g_1(\num{\bm{x}}), \dots, g_l(\num{\bm{x}})) \fmeq \num{y}.
                  \]
            \item[Minimization]
                  Again, by the inductive hypothesis $g$ is already represented, permitting this representation of $f$:
                  \[
                    \rep{f}(\bm{\qu{x}}, \fm{y}) \enskip\coloneqq\enskip g(\bm{\qu{x}}, \fm{y}) \fmeq \fm{0} \land \forall \fm{z} < \fm{y} \colon \neg g(\bm{\qu{x}}, \fm{z}) \fmeq \fm{0}
                  \]
                  Suppose $f(\bm{x}) = y$. Then $g(\bm{x}, y) = 0$ and $g(\bm{x}, z) \neq 0$ for all $z < y$.
                  \begin{description}
                    \item[$\Q \vdash \rep{f}(\num{\bm{x}}, \num{y})$]
                          The first conjunct $g(\num{\bm{x}}, \num{y}) \fmeq \fm{0}$ easily follows by the \hyperref[lm-run]{Runtime Lemma}. And using the \hyperref[lm-bounded]{Bounded Quantification Lemma} we can turn the second conjunct into a finite number of cases which can then be treated similarly. (If $y = 0$, \hyperref[lm-noneg]{non-negativity} can be used instead.)
                    \item[$\Q \vdash \exists! \fm{y} \colon \rep{f}(\num{\bm{x}}, \fm{y})$]
                          Existence has already been handled. And for uniqueness let $\fmc{y}$ be an arbitrary object such that $g(\num{\bm{x}}, \fmc{y}) \fmeq \fm{0} \land \forall \fm{z} < \fmc{y} \colon \neg g(\num{\bm{x}}, \fm{z}) \fmeq \fm{0}$. Then together with $\rep{f}(\num{\bm{x}}, \num{y})$ we can easily rule out $\fmc{y} < \num{y}$ and $\num{y} < \fmc{y}$ which by the \hyperref[lm-total-num]{totality of ordering} leaves only $\fmc{y} = \num{y}$.\qedhere
                  \end{description}
          \end{description}
  \end{description}
\end{prf}

\section{Diagonalization}

With the technical groundwork laid, we can now proceed to the main event. As already pointed out, the representation of a computable function is simply its source code in formal language. Moreover, it is a standard trick in computer science to achieve diagonalization by running a program on itself. So then, let's run a program on its own representation and see what happens.

\begin{definition}
  Writing $A(\rep{f})$ in the signature of an algorithm $A$ means that $A$ expects a representation as input, akin to pattern matching. For invalid inputs (which we'll avoid) the algorithm is allowed to arbitrarily misbehave.
\end{definition}

\begin{lemma}[Diagonal Lemma]\label{lm-diag}
  Let $\T$ be a theory which represents every computable function. And let $\qu{P}$ be a formula with one free variable. Then there exists a sentence $\fm{A}$ referred to as a ``fixed point'' such that $\T \vdash \fm{A} \leftrightarrow \fm{P}(\num{\fm{A}})$.
\end{lemma}

\begin{prf}
  Consider the following algorithm.
  \begin{algo}
    \Function{$D(\rep{f})$}{}
    \State\Return $\fm{P}(f(\numrep{f}))$
    \EndFunction
  \end{algo}
  Then $\T \vdash D(\numrep{D}) \fmeq \num{\fm{P}(D(\numrep{D}))\!}$ by the \hyperref[lm-run]{Runtime Lemma} and hence $\T \vdash \fm{P}(D(\numrep{D})) \leftrightarrow \fm{P}(\num{\fm{P}(D(\numrep{D}))\!})$ by \hyperref[lm-leibniz]{Leibniz's Law}, giving fixed point $\fm{P}(D(\numrep{D}))$.
\end{prf}

The main use of the Diagonal Lemma is to produce the Gödel sentence. Consider the following predicate.
\begin{definition}[$\Pvb_\fm{G}$]
  Let $\T$ be a computably enumerable theory which represents every computable function. And let $\pc$ be a proof verifier for $\T$. Then
  \[
    \Pvb_\fm{G}(\fm{x}) \enskip\coloneqq\enskip \exists \fm{p} \colon \pc(\fm{p}, \fm{x}) \fmeq \num{\cmark}.
  \]
\end{definition}
If one applies the Diagonal Lemma to $\neg \Pvb_\fm{G}$, one produces a sentence $\fm{G}$ such that $\PA$ proves $\fm{G} \leftrightarrow \neg\Pvb_\fm{G}(\num{\fm{G}})$. Informally this sentence states: ``I am not provable.'' which, assuming that $\PA$ is consistent, turns out to be true. Furthermore, one can also show that $\neg \fm{G}$ can't be provable either. Though this needs the stronger assumption that $\PA$ is \emph{correct} (also referred to as \emph{sound}); if $\PA$ proves something, then it is actually so. Or to be pedantic it at least requires $\omega$-consistency; if $\PA$ proves that some unbounded search terminates, then it actually does. So in summary, if $\PA$ is $\omega$-consistent, it is incomplete.

For the audience in 1931 this already settled the matter. Still, in 1936 John B. Rosser found a way to iron out the last wrinkle as well by using this predicate instead \cite{rosser}:
\begin{definition}[Concatenation]
  $\cat$ is the computable function which concatenates two numbers. I.e., it returns the number which is written using the digits of the first number followed by the digits of the second number. Furthermore, we'll use left-associative infix notation for this function, so $x \cat y \cat z \coloneqq (x \cat y) \cat z$. The fact that this function is associative won't be needed.
\end{definition}

\begin{definition}[$\Pvb_{\mathsf{R}}$]
  Let $\T$ be a computably enumerable extension of $\Q$. And let $\pc$ be a proof verifier for $\T$. Then
  \[
    \Pvb_\fm{R}(\fm{x}) \enskip\coloneqq\enskip \exists \fm{p} \colon \pc(\fm{p}, \fm{x}) \fmeq \num{\cmark} \land \forall \fm{q} \leq \fm{p} \colon \neg\, \pc(\fm{q}, \num{\neg} \cat \fm{x}) \fmeq \num{\cmark}.
  \]
\end{definition}
If one applies the Diagonal Lemma to $\neg\Pvb_\fm{R}$, it produces the Rosser sentence $\fm{R}$ which can be paraphrased as: ``For every proof of me there exists a shorter disproof.'' This sentence, unlike $\fm{G}$, can be demonstrated to be neither provable nor disprovable assuming only the consistency of $\PA$, giving us the much nicer formulation: if $\PA$ is consistent, it is incomplete.

Rosser's argument comes with the minor drawback that just to have $\leq$ available one already has to narrow the discussion from theories which represent every computable function to just extensions of $\Q$. And one also still has spell out a few more details. For starters, one has to invoke the Bounded Quantification Lemma. This brings us then to another way of proving the First Incompleteness Theorem which in my mind fully captures the essence with beautiful concision.
\begin{theorem}[First Incompleteness Theorem]\label{thm-first}
  Let $\T$ be a consistent and computably enumerable theory which represents every computable function. Then $\T$ is incomplete.
\end{theorem}

\begin{prf}
  First, we construct the following algorithm.
  \begin{algo}
    \Function{$X(\rep{f})$}{}
    \For{$p \in \mathbb{N}$}
    \If{$p$ proves $\neg\, f(\numrep{f}) \fmeq \num{\cmark}$}
    \State\Return \hspace{-1.8pt} $\cmark$
    \EndIf
    \vspace{-2.75pt}
    \If{$p$ proves $\hphantom{\neg\,}f(\numrep{f}) \fmeq \num{\cmark}$}
    \State\Return \hspace{-1.8pt} $\xmark$
    \EndIf
    \EndFor
    \EndFunction
  \end{algo}
  This lets us produce
  \[
    \fm{R} \enskip\coloneqq\enskip \neg\, X(\numrep{X}) \fmeq \num{\cmark}.
  \]
  Now suppose that $\fm{R}$ was decidable. Then there must exist a \emph{first} $p \in \mathbb{N}$ which decides $\fm{R}$. This leads to two cases.
  \newline
  \begin{minipage}{0.46\textwidth}
    \smallskip
    \begin{description}[leftmargin=0.5cm]
      \item[$p$ proves $\mathsf{R}$]
            Then $X(\rep{X})$ returns $\cmark$ upon finding $p$ and so $\T \vdash \neg \mathsf{R}$ by the \hyperref[lm-run]{Runtime Lemma}; an inconsistency.
    \end{description}
  \end{minipage}
  \hspace{0.05\textwidth}
  \begin{minipage}{0.48\textwidth}
    \smallskip
    \begin{description}[leftmargin=0.5cm]
      \item[$p$ proves $\neg \mathsf{R}$]
            Then $X(\rep{X})$ returns $\xmark$ upon finding $p$ and so $\T \vdash \mathsf{R}$ by the \hyperref[lm-run]{Runtime Lemma}; an inconsistency.\qedhere
    \end{description}
  \end{minipage}
\end{prf}

The sentence $\fm{R}$ used in the above proof carries its name for a reason. It's because, as some reflection will reveal, $\fm{R}$ is in fact a Rosser sentence, claiming: ``For every proof of me there exists a shorter disproof.'' Furthermore, in order to recover the original ``I am not provable'' one only needs to drop the last two lines of $X$.

\section{Re-Representation}

At a high level the Second Incompleteness Theorem isn't all that hard to grasp. We've already seen that if $\PA$ is consistent, then $\fm{G}$ isn't provable and thereby true. In short, $\Con \rightarrow \fm{G}$ where $\Con$ expresses the consistency of $\PA$. What's more, as Gödel first noted in 1931, this fact can be proved in $\PA$ as well since it can step through the proof of the First Incompleteness Theorem just the same. To coin a phrase, $\PA$ represents that it represents---or \emph{re-represents}---its own proof verifier. But that means that if $\PA$ could prove $\Con$, then $\PA$ could also prove $\fm{G}$ by modus ponens which is impossible. That's the Second Incompleteness Theorem: if $\PA$ is consistent, then it can't prove this to be so.

Alas, the crucial fact that $\Con \rightarrow \fm{G}$ can be proved in $\PA$ is notoriously difficult to establish rigorously. Gödel himself famously concluded his 1931 paper stating: ``The results will be stated and proved in fuller generality in a forthcoming sequel. There too, the mere outline proof we have given of [the Second Incompleteness Theorem] will be presented in detail.'' Such a sequel never came. Instead, the first rigorous demonstration appeared in 1939 in the second volume of \textit{Foundations of Mathematics} \cite{bernays-hilbert} written by David Hilbert and his assistant Paul Bernays---Bernays being the main author.

But besides technical issues, there are also conceptual ones. First and foremost, what exact expression should we choose for $\Con$? Gödel had the obvious in mind:
\begin{definition}[$\Con_\fm{G}$]
  \hspace{2.25cm}$\Con_\fm{G} \enskip\coloneqq\enskip \neg \exists \fm{x} \colon \Pvb_\fm{G}(\fm{x}) \land \Pvb_\fm{G}(\num{\neg} \cat \fm{x})$
\end{definition}
However, Gödel had many freedoms when making his construction. What made the choice he arrived at the right one? This question is brought to a head when one examines the alternative derived from Rosser's predicate.
\begin{definition}[$\Con_\fm{R}$]
  \hspace{2.25cm}$\Con_\fm{R} \enskip\coloneqq\enskip \neg \exists \fm{x} \colon \Pvb_\fm{R}(\fm{x}) \land \Pvb_\fm{R}(\num{\neg} \cat \fm{x})$
\end{definition}
If one were to use this consistency statement instead, then $\PA$ actually \emph{could} prove its own consistency. This happens for the simple reason that \emph{every} theory, whether consistent or not, is ``consistent'' according to the Rosser definition. To address this conundrum Hilbert and Bernays decided to put down three conditions which after some modification by Martin Löb became these:

\begin{definition}[Derivability Conditions]
  A predicate $\Pvb$ obeys the \emph{derivability conditions} for a theory $\T$ if the following holds for any sentences $\fm{A}, \fm{B}$:
  \begin{description}
    \item[Representability]
          If $\T \vdash \fm{A}$, then $\T \vdash \Pvb(\num{\fm{A}})$.
    \item[Distributivity]
          $\T \vdash \Pvb(\num{\fm{A} \rightarrow \fm{B}}) \rightarrow (\Pvb(\num{\fm{A}}) \rightarrow \Pvb(\num{\fm{B}}))$.
    \item[Re-Representability]
          $\T \vdash \Pvb(\num{\fm{A}}) \rightarrow \Pvb(\num{\Pvb(\num{\fm{A}})})$.
  \end{description}
\end{definition}

From these three conditions---also known as the HBL-derivability conditions (Hilbert, Bernays, Löb)---the Second Incompleteness Theorem can be derived in short order. This both shuts the door on Rosser's predicate and provides a helpful layer of abstraction, similar to the representability of computable functions.

The only question left on the table is then: does some proposed provability predicate such as $\Pvb_\fm{G}$ actually meet the criteria? This is where all the fearsome details end up hiding; details which crucially rely upon the exact representation of the proof verifier $\pc$ underlying $\Pvb_\fm{G}$, lest any Rosserian shenenigans take place under the hood. The standard approach runs roughly as follows.
\begin{description}
  \item[Representability]
        This part carries over from studying the First Incompleteness Theorem and is basically free.
  \item[Distributivity]
        Here one can show that $\PA$ recognizes the modus ponens inference rule in $\Pvb_\fm{G}$. That is to say, $\PA$ can prove a concatenation of a proof of $\fm{A} \rightarrow \fm{B}$, a proof of $\fm{A}$, and the list containing only $\fm{B}$ to be a proof of $\fm{B}$. Doing so requires some cursory inspection of $\Pvb_\fm{G}$ as well as wading through a fair number of details involving elementary features of lists. But it's still manageable.
  \item[Re-Representability]
        This is the true villain of the story. In outline the strategy is to show that
        \[
          \PA \vdash \fm{A} \rightarrow \Pvb_\fm{G}(\num{\fm{A}})
        \]
        for any sentence $\fm{A}$ built without unbounded universal quantifiers. Re-representability is then obtained as a special case. But checking the above demands an exceedingly elaborate structural induction on the construction of $\fm{A}$, during each stage of which one has to x-ray $\Pvb_\fm{G}$. And if that wasn't enough, realize that as soon as one moves on from $\PA$ to any of its extensions one is presented with a new provability predicate for that theory and one has to start all over again.
\end{description}

The challenge ahead of us is daunting. So instead of charging straight ahead let's pause for a moment and try to figure out what is fundamentally causing the difficulty. It's that, whereas the modus ponens inference rule is usually coded into $\Pvb_\fm{G}$ at a high level, re-representability emerges piece by piece from the entirety of the construction. But picking $\Pvb_\fm{G}$ as our provability predicate was a deliberate choice; a choice which we can amend. If only we could find a provability predicate which carried the re-representability property at the surface, then we might save ourselves a lot of grief during the verification.

Indeed, why not simply pick the following provability predicate and declare victory?
\[
  \Pvb_\top(\fm{x}) \enskip\coloneqq\enskip \fm{x} \fmeq \fm{x}
\]
This predicate trivially satisfies all three conditions. Yet something is clearly amiss here. According to $\Pvb_\top$ everything is provable. In other words, $\Pvb_\top$ encodes the belief that our theory is inconsistent. Yes, the Second Incompleteness Theorem still applies to this predicate. But it has degenerated into the trite observation that a consistent theory which believes itself to be inconsistent can't simultaneously assert its own consistency. Evidently we're still missing a basic requirement.

\begin{definition}[Provability Predicate]
  A predicate $\Pvb$ is considered a \emph{provability predicate} for a theory $\T$ if it obeys the following condition for any sentence $\fm{A}$.
  \begin{description}
    \item[Correctness] $\Pvb(\num{\fm{A}})$ is true (in the intended interpretation) if and only if $\T \vdash \fm{A}$.
  \end{description}
\end{definition}

I've seen others make the alternative mandate that if $\T \vdash \Pvb(\num{\fm{A}})$, then $\T \vdash \fm{A}$. But this ends up roping in the assertion that $\T$ is a correct theory and would conflict with our aim to rest the Second Incompleteness Theorem on consistency alone.

\begin{definition}[Weak Provability Predicate]
  A predicate $\Pvb$ is considered a \emph{weak} or \emph{representable provability predicate} for a theory $\T$ if it is a provability predicate for $\T$ which also obeys the following condition for any sentence $\fm{A}$.
  \begin{description}
    \item[Representability] If $\T \vdash \fm{A}$, then $\T \vdash \Pvb(\num{\fm{A}})$.
  \end{description}
\end{definition}

\begin{definition}[Strong Provability Predicate]
  A predicate $\Pvb$ is considered a \emph{strong provability predicate} for a theory $\T$ if it is a provability predicate for $\T$ which also obeys the derivability conditions. To repeat, for any sentences $\fm{A}, \fm{B}$:
  \begin{description}
    \item[Representability]
          If $\T \vdash \fm{A}$, then $\T \vdash \Pvb(\num{\fm{A}})$.
    \item[Distributivity]
          $\T \vdash \Pvb(\num{\fm{A} \rightarrow \fm{B}}) \rightarrow (\Pvb(\num{\fm{A}}) \rightarrow \Pvb(\num{\fm{B}}))$.
    \item[Re-Representability]
          $\T \vdash \Pvb(\num{\fm{A}}) \rightarrow \Pvb(\num{\Pvb(\num{\fm{A}})})$.
  \end{description}
\end{definition}

Assembling a predicate with the correctness property all the way from scratch would take a while. So let me set up an intermediate abstraction to build on top of.

\begin{definition}[Correctly Representable Function]
  A function $f$ is considered to be \textit{correctly represented} in a theory $\T$ if it is represented in $\T$ by a formula $\rep{f}$ and $\rep{f}(\num{\bm{x}}, \num{y})$ is true if and only if $f(\bm{x}) = y$.
\end{definition}
The following should then come with little surprise.
\begin{theorem}
  $\Q$ correctly represents every computable function.
\end{theorem}

\begin{prf-sketch}
All that's required for this result is to retrace the argument by which \hyperref[thm-qrep]{$\Q$ represents every computable function} and check that correctness is satisfied at each step.
\end{prf-sketch}

It's worth bearing in mind that the ability of a theory $\T$ to correctly represent a function by no means implies that $\T$ is a correct theory. For instance, since $\Q$ correctly represents every computable function so does every extension of it, including the inconsistent ones.

We can then immediately check off the first two boxes by noting that $\Pvb_\fm{G}$ (and $\Pvb_\fm{R}$ provided our theory is consistent) is a weak provability predicate as long as it has been constructed with a correct representation of $\pc$. But now for the heart of the matter: can we find a \emph{strong} provability predicate which is also easily recognized as such? Actually, we can, starting with the following idea.

Take a closer look at this example consequent of the re-representability condition: $\Pvb(\num{\Pvb(\num{\fm{S}\fm{0} + \fm{S}\fm{0} \fmeq \fm{S}\fm{S}\fm{0}})})$. Antropomorphizing our provability predicate lets us rephrase this in English as: ``I believe that I believe that $1 + 1 = 2$.'' Now answer me this: if I ever asked you if you believed this statement yourself, would you then dutifully proceed to simulate your own brain as it is pondering whether $1 + 1 = 2$? I doubt it. More likely you'd just ponder the question directly. But then why not permit $\Pvb$ the same liberty? Or more generally, why couldn't $\fm{A}$ be used to derive $\Pvb(\num{\fm{A}})$ straight away? In customary notation we'll then have these two inference rules built into our provability predicate:
\[
  \text{Modus Ponens:}\quad\frac{\fm{A} \rightarrow \fm{B} \hspace{16pt} \fm{A}}{\fm{B}} \hspace{3cm} \text{Representability:}\quad\frac{\fm{A}}{\Pvb(\num{\fm{A}})}
\]
We're going to have to meet the representability condition---if $\fm{A}$ is a theorem, so is $\Pvb(\num{\fm{A}})$---anyway. So this is a completely sound inference to make. The only puzzle this leaves us with is to find a way for $\Pvb$ to recognize itself. But at this point that's a well-practiced routine. The solution, as always, is to run a program on its own source code.

\begin{lemma}[Totality Of Ordering]\label{lm-total}
  \[
    \PA \vdash \forall \fm{x}, \fm{y} \colon \fm{x} < \fm{y} \lor \fm{y} < \fm{x} \lor \fm{x} \fmeq \fm{y}.
  \]
\end{lemma}

\begin{prf-sketch}
Repeat the proof for the \hyperref[lm-total-num]{totality of ordering w.r.t. numerals} while carrying out the induction within $\PA$.
\end{prf-sketch}

\begin{definition}[List]
  A \emph{list} is a single number which contains a finite ordered set of other numbers via some coding. Furthermore, we'll write $[x_1, \dots, x_k]$ to denote the list whose items are $x_1, \dots, x_k$. Strictly speaking, this is an infinite set of computable functions, one for each number of inputs. So there are $[\cdot], [\cdot, \cdot]$, $[\cdot, \cdot, \cdot]$, etc.
\end{definition}

\begin{definition}[List Access]
  Further extending our use of computer programming notation we'll use $\cdot[\cdot]$ to denote the computable function which given a list $[\bm{x}]$ and index $i$ retrieves the corresponding item, namely $x_i$. In short: $[\bm{x}][i] = x_i$.
\end{definition}

\begin{definition}[List Length]
  $\lVert \cdot \rVert$ denotes the computable function which determines the length of a list.
\end{definition}

\begin{definition}[List Concatenation]
  $\Cat$ denotes the computable function which concatenates two lists. And analogously to basic concatenation we'll use left-associative infix notation for this function.
\end{definition}

\begin{lemma}[Basic Properties Of Lists]\label{lm-list}
  For some correct representations of $\cdot[\cdot]$, $\lVert \cdot \rVert$, and $\Cat$ we have:
  \newline
  \begin{minipage}{0.4\textwidth}
    \begin{align*}
       & \PA \vdash \forall \fm{x}, \fm{i} \colon \exists \fm{y} \colon \fm{x}[\fm{i}] \fmeq \fm{y}, \tag{L0} \label{lm-list.0}     \\
       & \PA \vdash \forall \fm{x} \colon \exists \fm{n} \colon \lVert \fm{x} \rVert = \fm{n}, \tag{L1} \label{lm-list.1}           \\
       & \PA \vdash \forall \fm{x}, \fm{y} \colon \exists \fm{z} \colon \fm{x} \Cat \fm{y} \fmeq \fm{z}, \tag{L2} \label{lm-list.2}
    \end{align*}
  \end{minipage}
  \begin{minipage}{0.5\textwidth}
    \begin{align*}
       & \PA \vdash \forall \fm{x}, \fm{y} \colon \forall \fm{i} < \lVert \fm{x} \rVert \colon \fm{x}[\fm{i}] \fmeq (\fm{x} \Cat \fm{y})[\fm{i}] , \tag{L3} \label{lm-list.3}                       \\
       & \PA \vdash \forall \fm{x}, \fm{y} \colon \forall \fm{i} < \lVert \fm{y} \rVert \colon \fm{y}[\fm{i}] \fmeq (\fm{x} \Cat \fm{y})[\lVert \fm{x} \rVert + \fm{i}], \tag{L4} \label{lm-list.4} \\
       & \PA \vdash \forall \fm{x}, \fm{y} \colon \lVert \fm{x} \Cat \fm{y} \rVert \fmeq \lVert \fm{x} \rVert + \lVert \fm{y} \rVert. \tag{L5} \label{lm-list.5}
    \end{align*}
  \end{minipage}
\end{lemma}

\begin{prf}
  Omitted.
\end{prf}

The preceding lemma is already baked into the definition of concatenation in Gödel's original construction. Or at least the right half of it is. Perhaps that's also the reason why Hilbert and Bernays left out further details when they listed the right half on page 321 of their second volume. Also, L0 is never used. I only added it to complete the set.

\begin{theorem}[Existence Theorem For Strong Provability Predicates]\label{thm-existence}
  Let $\T$ be a computably enumerable extension of $\PA$. Then there exists a strong provability predicate for $\T$.
\end{theorem}

\begin{prf}
  We start with an arbitrary predicate which we know to be a weak provability predicate. $\Pvb_\fm{G}$ will do nicely. We can then bolt on the distributivity and re-representability properties by first constructing the following algorithm.
  \begin{algo}
    \Function{$Q(\rep{f}, x)$}{}
    \State\Return $\begin{aligned}[t]
        \exists \fm{p}, \fm{m} \colon & \fm{p}[\fm{m}] \fmeq \num{x} \, \land            \\
                                      & \lVert \fm{p} \rVert \fmeq \fm{S}\fm{m} \, \land \\
                                      & \forall \fm{i} < \lVert \fm{p} \rVert \colon
        \begin{aligned}[t]
           & \PvbG(\fm{p}[\fm{i}]) \, \lor          \\
           & \exists \fm{j}, \fm{k} < \fm{i} \colon
          \begin{aligned}[t]
             & \fm{p}[\fm{j}] \fmeq \fm{p}[\fm{k}] \cat \num{\rightarrow} \cat \fm{p}[\fm{i}] \, \lor \\
             & \fm{p}[\fm{i}] \fmeq f(\numrep{f}, \fm{p}[\fm{j}])
          \end{aligned}
        \end{aligned}
      \end{aligned}$
    \EndFunction
  \end{algo}
  And then we have the accompanying provability predicate:
  \[
    \Pvb(\fm{x}) \enskip\coloneqq\enskip
    \begin{aligned}[t]
      \exists \fm{p}, \fm{m} \colon & \fm{p}[\fm{m}] \fmeq \fm{x} \, \land             \\
                                    & \lVert \fm{p} \rVert \fmeq \fm{S}\fm{m} \, \land \\
                                    & \forall \fm{i} < \lVert \fm{p} \rVert \colon
      \begin{aligned}[t]
         & \PvbG(\fm{p}[\fm{i}]) \, \lor          \\
         & \exists \fm{j}, \fm{k} < \fm{i} \colon
        \begin{aligned}[t]
           & \fm{p}[\fm{j}] \fmeq \fm{p}[\fm{k}] \cat \num{\rightarrow} \cat \fm{p}[\fm{i}] \, \lor \\
           & \fm{p}[\fm{i}] \fmeq Q(\numrep{Q}, \fm{p}[\fm{j}])
        \end{aligned}
      \end{aligned}
    \end{aligned}\tag*{\parbox[t]{6cm}{\raggedleft\footnotesize (there exists a proof $\fm{p}$ containing $\fm{x}$)                                                                               \\[5pt] ($\fm{x}$ is the last item)\\[5pt] (every item of $\fm{p}$ is either $\PvbG$-provable)\\[5pt] (or follows by modus ponens)\\[5pt] (or follows by representability)}}
  \]

  Onwards to the verfication. In what follows we'll be using \hyperref[lm-leibniz]{Leibniz's Law for representable terms}, the \hyperref[lm-run]{Runtime Lemma}, and the various \hyperref[lm-list]{properties of lists} all over the place. I've omitted references for brevity. Additionally, we'll postpone correctness to the end.
  \begin{description}
    \item[Representability]
          If $\T \vdash \fm{A}$, then $[\fm{A}]$ by itself already constitutes a ``proof'' and we're naturally led to the claim
          \[
            \num{[\fm{A}]}[\fm{0}] = \num{\fm{A}} \land \lVert \num{\fm{[A]}} \rVert = \fm{S}\fm{0} \land \forall \fm{i} < \lVert \num{[\fm{A}]} \rVert \colon \PvbG(\num{[\fm{A}]}[\fm{i}])
          \]
          which can be seen to be provable through the \hyperref[lm-bounded]{Bounded Quantification Lemma} and the representability property of $\PvbG$. The desired conclusion $\Pvb(\num{\fm{A}})$ follows without effort.
    \item[Distributivity]
          For this case we need to establish that $\T$ can argue from the assumptions $\Pvb(\num{\fm{A} \rightarrow \fm{B}})$ and $\Pvb(\num{\fm{A}})$ to the conclusion $\Pvb(\num{\fm{B}})$. More concretely, we'll want to show that if we instantiate the two existentials $\Pvb(\num{\fm{A} \rightarrow \fm{B}})$ and $\Pvb(\num{\fm{A}})$ with $\fmc{p}_{\fm{A} \imp \fm{B}}$ and $\fmc{p}_\fm{A}$ respectively, then we can obtain $\fmc{p}_{\fm{B}\fmtick}, \fmc{p}_\fm{B}$, and $\fmc{m}_\fm{B}$ such that
          \[
            \fmc{p}_{\fm{B}\fmtick} \fmeq \fmc{p}_{\fm{A} \imp \fm{B}} \Cat \fmc{p}_\fm{A}, \quad \fmc{p}_\fm{B} \fmeq \fmc{p}_{\fm{B}\fmtick} \Cat \num{[\fm{B}]}, \quad\text{and}\quad \fmc{m}_\fm{B} \fmeq \lVert \fmc{p}_{\fm{B}\fmtick} \rVert
          \]
          and that $\fmc{p}_\fm{B}, \fmc{m}_\fm{B}$ will then be able to serve as witnesses for $\Pvb(\num{\fm{B}})$.

          To that end we can immediately get the assertion that $\num{\fm{B}}$ is the last item of $\fmc{p}_\fm{B}$ out of the way by observing that
          \[
            \fmc{p}_\fm{B}[\fmc{m}_\fm{B}] \fmeq (\fmc{p}_{\fm{B}\fmtick} \Cat \num{[\fm{B}]})[\lVert \fmc{p}_{\fm{B}\fmtick} \rVert + \fm{0}] \fmeq \num{[\fm{B}]}[\fm{0}] \fmeq \num{\fm{B}} \qquad\text{and}\qquad \lVert \fmc{p}_\fm{B} \rVert \fmeq \lVert \fmc{p}_{\fm{B}\fmtick} \rVert + \lVert \num{[\fm{B}]} \rVert \fmeq \fm{S}\fmc{m}_\fm{B}.
          \]

          What's left then is to verify that every item of $\fmc{p}_\qu{B}$ is valid, i.e.:
          \[
            \forall \fm{i} < \lVert \fmc{p}_\fm{B} \rVert \colon \PvbG(\fmc{p}_\fm{B}[\fm{i}]) \;\lor\; \exists \fm{j}, \fm{k} < \fm{i} \colon\; \fmc{p}_\fm{B}[\fm{j}] \fmeq \fmc{p}_\fm{B}[\fm{k}] \cat \num{\rightarrow} \cat \fmc{p}_\fm{B}[\fm{i}] \;\lor\; \fmc{p}_\fm{B}[\fm{i}] \fmeq Q(\numrep{Q}, \fmc{p}_\fm{B}[\fm{j}]).
          \]
          This can be established for an arbitrary $\fmc{i}$ such that $\fmc{i} < \lVert \fmc{p}_\fm{B} \rVert$ by using the \hyperref[lm-total]{totality of ordering} to break the argument down into three cases.
          \begin{description}
            \item[$\fmc{i} < \lVert \fmc{p}_{\fm{A} \imp \fm{B}} \rVert$]
                  Here we can carry out the following conversion:
                  \[
                    \begin{aligned}
                       & \PvbG(\fmc{p}_{\fm{A} \imp \fm{B}}[\fmc{i}]) \, \lor \\
                       & \exists \fm{j}, \fm{k} < \fmc{i} \colon
                      \begin{aligned}[t]
                         & \fmc{p}_{\fm{A} \imp \fm{B}}[\fm{j}] \fmeq \fmc{p}_{\fm{A} \imp \fm{B}}[\fm{k}] \cat \num{\rightarrow} \cat \fmc{p}_{\fm{A} \imp \fm{B}}[\fmc{i}] \, \lor \\
                         & \fmc{p}_{\fm{A} \imp \fm{B}}[\fmc{i}] \fmeq Q(\numrep{Q}, \fmc{p}_{\fm{A} \imp \fm{B}}[\fm{j}])
                      \end{aligned}
                    \end{aligned}
                    \qquad\leftrightarrow\qquad
                    \begin{aligned}
                       & \PvbG(\fmc{p}_\fm{B}[\fmc{i}]) \, \lor  \\
                       & \exists \fm{j}, \fm{k} < \fmc{i} \colon
                      \begin{aligned}[t]
                         & \fmc{p}_\fm{B}[\fm{j}] \fmeq \fmc{p}_\fm{B}[\fm{k}] \cat \num{\rightarrow} \cat \fmc{p}_\fm{B}[\fmc{i}] \, \lor \\
                         & \fmc{p}_\fm{B}[\fmc{i}] \fmeq Q(\numrep{Q}, \fmc{p}_\fm{B}[\fm{j}])
                      \end{aligned}
                    \end{aligned}
                  \]
                  Of course having obtained the left-hand side from $\Pvb(\num{\fm{A} \rightarrow \fm{B}})$. Let's consider an example to illustrate; the substitution of $\fmc{p}_{\fm{A} \imp \fm{B}}[\fm{k}]$ on the left with $\fmc{p}_\fm{B}[\fm{k}]$ on the right. Suppose we had instantiated $\fm{k}$ on the left with $\fmc{k}_{\fm{A} \imp \fm{B}}$. Then since $\fmc{k}_{\fm{A} \imp \fm{B}} < \fmc{i} < \lVert \fmc{p}_{\fm{A} \imp \fm{B}} \rVert < \lVert \fmc{p}_{\fm{B}\fmtick} \rVert$ we can deduce
                  \[
                    \fmc{p}_{\fm{A} \imp \fm{B}}[\fmc{k}_{\fm{A} \imp \fm{B}}] \fmeq (\fmc{p}_{\fm{A} \imp \fm{B}} \Cat \fmc{p}_\fm{A})[\fmc{k}_{\fm{A} \imp \fm{B}}] \fmeq \fmc{p}_{\fm{B}\fmtick}[\fmc{k}_{\fm{A} \imp \fm{B}}] \fmeq (\fmc{p}_{\fm{B}\fmtick} \Cat \num{[\fm{B}]})[\fmc{k}_{\fm{A} \imp \fm{B}}] \fmeq \fmc{p}_\fm{B}[\fmc{k}_{\fm{A} \imp \fm{B}}].
                  \]
                  This permits the substitution we've aimed for. We then simply repeat this process for all the other representable terms and finally reintroduce existential quantifiers to arrive at the right-hand side.
            \item[$\lVert \fmc{p}_{\fm{A} \imp \fm{B}} \rVert \leq \fmc{i} < \lVert \fmc{p}_{\fm{B}\fmtick} \rVert$]
                  Here we evidently have $\fmc{i} = \lVert \fmc{p}_{\fm{A} \imp \fm{B}} \rVert + \fmc{i}_\fm{A}$ for some $\fmc{i}_\fm{A} < \lVert \fmc{p}_\fm{A} \rVert$ which allows us to proceed the same way we did in the previous case, this time lifting over the items of $\fmc{p}_\fm{A}$.
            \item[$\fmc{i} \fmeq \lVert \fmc{p}_{\fm{B}\fmtick} \rVert$]
                  Finally, for the last item we can use $\Pvb(\num{\fm{A} \rightarrow \fm{B}})$ and $\Pvb(\num{\fm{A}})$ to obtain $\fmc{m}_\qu{\fm{A} \imp \fm{B}}$ and $\fmc{m}_\fm{A}$ such that
                  \[
                    \fmc{p}_{\fm{A} \imp \fm{B}}[\fmc{m}_{\fm{A} \imp \fm{B}}] \fmeq \num{\fm{A} \rightarrow \fm{B}},\quad \lVert \fmc{p}_{\fm{A} \imp \fm{B}} \rVert \fmeq \fm{S}\fmc{m}_{\fm{A} \imp \fm{B}}, \quad \fmc{p}_\fm{A}[\fmc{m}_\fm{A}] \fmeq \num{\fm{A}}, \quad\text{and}\quad \lVert \fmc{p}_\fm{A} \rVert \fmeq \fm{S}\fmc{m}_\fm{A}.
                  \]
                  Thus, after obtaining $\fmc{m}_{\fm{B}\fmtick}$ such that $\fmc{m}_{\fm{B}\fmtick} \fmeq \lVert \fmc{p}_{\fm{A} \imp \fm{B}} \rVert + \fmc{m}_\fm{A}$ we can gather together
                  \[
                    \fmc{p}_\fm{B}[\fmc{m}_{\fm{A} \imp \fm{B}}] \fmeq \fmc{p}_{\fm{A} \imp \fm{B}}[\fmc{m}_{\fm{A} \imp \fm{B}}] \fmeq \num{\fm{A} \rightarrow \fm{B}}, \quad \fmc{p}_\fm{B}[\fmc{m}_{\fm{B}\fmtick}] \fmeq \fmc{p}_\fm{A}[\fmc{m}_\fm{A}] \fmeq \num{\fm{A}}, \quad\text{and}\quad \fmc{p}_\fm{B}[\fmc{i}] \fmeq \fmc{p}_\fm{B}[\fmc{m}_\fm{B}] \fmeq \num{\fm{B}}.
                  \]
                  And so in all we get
                  \[
                    \fmc{p}_\fm{B}[\fmc{m}_{\fm{A} \imp \fm{B}}] \fmeq \num{\fm{A} \rightarrow \fm{B}} \fmeq \num{\fm{A}} \cat \num{\rightarrow} \cat \num{\fm{B}} \fmeq \fmc{p}_\fm{B}[\fmc{m}_{\fm{B}\fmtick}] \cat \num{\rightarrow} \cat \fmc{p}_\fm{B}[\fmc{i}].
                  \]
                  Thus, noting $\fmc{m}_{\fm{A} \imp \fm{B}} < \fmc{m}_{\fm{B}\fmtick} < \fmc{i}$, we find
                  \[
                    \exists \fm{j}, \fm{k} < \fmc{i} \colon \fm{p}_\fm{B}[\fm{j}] \fmeq \fm{p}_\fm{B}[\fm{k}] \cat \num{\rightarrow} \cat \fm{p}_\fm{B}[\fmc{i}]
                  \]
                  which satisfies the consequent.
          \end{description}
    \item[Re-Representability]
          Here we need to guide $\T$ from the assumption $\Pvb(\num{\fm{A}})$ to the conclusion $\Pvb(\num{\Pvb(\num{\fm{A}})})$. With our new inference rule in place this is now as simple as observing that
          \[
            \fmc{p}_{\fmc{p}_\fm{A}} \fmeq \fmc{p}_\fm{A} \Cat \num{[\Pvb(\num{\fm{A}})]},
          \]
          where $\fmc{p}_\fm{A}$ has been obtained from $\Pvb(\num{\fm{A}})$, serves as the desired proof. The verification proceeds along exactly the same lines as it did for the distributive property, so we'll skip most of it. The only novelty occurs when checking the final item at index $\fmc{i} \fmeq \lVert \fmc{p}_\fm{A} \rVert$. Here the assumption provides an $\fmc{m}_\fm{A}$ such that
          \[
            \fmc{p}_\fm{A}[\fmc{m}_\fm{A}] \fmeq \num{\fm{A}} \quad\text{and}\quad \lVert \fmc{p}_\fm{A} \rVert \fmeq \fm{S}\fmc{m}_\fm{A}.
          \]
          So because $\fmc{m}_\fm{A} < \lVert \fmc{p}_\fm{A} \rVert$ we get
          \[
            \fmc{p}_{\fmc{p}_\fm{A}}[\fmc{m}_\fm{A}] \fmeq \fmc{p}_\fm{A}[\fmc{m}_\fm{A}] \fmeq \num{\fm{A}}.
          \]
          Of course, we also have $\fmc{p}_{\fmc{p}_\fm{A}}[\lVert \fmc{p}_\fm{A} \rVert] \fmeq \num{\Pvb(\num{\fm{A}})}$. Therefore,
          \[
            \fmc{p}_{\fmc{p}_\fm{A}}[\fmc{i}] \fmeq \num{\Pvb(\num{\fm{A}})} \fmeq Q(\numrep{Q}, \num{\fm{A}}) \fmeq Q(\numrep{Q}, \fmc{p}_{\fmc{p}_\fm{A}}[\fmc{m}_\fm{A}])
          \]
          and that gives the sought after
          \[
            \exists \fm{j} < \fmc{i} \colon \fmc{p}_{\fmc{p}_\fm{A}}[\fmc{i}] \fmeq Q(\numrep{Q}, \fmc{p}_{\fmc{p}_\fm{A}}[\fm{j}]).
          \]
    \item[Correctness]
          Lastly, the correctness of $\Pvb$ follows from a straightforward inductive argument. In the forward direction, suppose $\Pvb(\num{\fm{A}})$ is true. Then there really does exist a proof $p$ with the outlined characteristics. In particular, we can see that every item of $p$---which includes $\fm{A}$---must be a theorem. By induction on the index $i$:
          \begin{description}
            \item[Base Case]
                  If $i = 0$, then we must have $\PvbG(\num{p[i]})$ which makes $p[i]$ provable by the correctness of $\PvbG$.
            \item[Induction]
                  Now let's assume that the items of $p$ up to index $i$ are known to be theorems. Then so is the $i+1$st item since $\Pvb(\num{\fm{A}})$ assures us that at least one of the following is true:
                  \begin{itemize}
                    \item The $i+1$st item is provable by the correctness of $\PvbG$.
                    \item The $i+1$st item follows by modus ponens from a pair of previous items which we know to be theorems by inductive assumption.
                    \item The $i+1$st item is of the form $\Pvb(\num{\fm{A}\hspace{-0.15em}\fmtick})$ for some preceding theorem $\fm{A}\hspace{-0.15em}\fmtick$, making the $i+1$st item a theorem by the already demonstrated representability property of $\Pvb$.
                  \end{itemize}
          \end{description}
          And in the other direction, if $\T \vdash \fm{A}$, then arguing as we did for representability---except relying on correctness instead of provability---we can see that $[\fm{A}]$ satisfies the existential claim expressed by $\Pvb(\num{\fm{A}})$, making it true.\qedhere
  \end{description}
\end{prf}

Phew! With that slog behind us let's take the scenic route past Löb's Theorem towards the Second Incompleteness Theorem.

\begin{theorem}[Löb's Theorem]\label{thm-loeb}
  Let $\Pvb$ be a predicate obeying the derivability conditions for some theory $\T$ which represents every computable function. Then the following holds for any sentence $\fm{A}$.
  \[
    \text{If } \T \vdash \Pvb(\num{\fm{A}}) \rightarrow \fm{A}, \text{ then } \T \vdash \fm{A}.
  \]
\end{theorem}

\begin{prf}
  We begin by using the \hyperref[lm-diag]{Diagonal Lemma} to construct a fixed point $\fm{L}$ of $\Pvb(\fm{x}) \rightarrow \fm{A}$ so that
  \[
    \T \vdash \fm{L} \leftrightarrow (\Pvb(\num{\fm{L}}) \rightarrow \fm{A}).
  \]
  We can then use the representability and distributivity conditions of $\Pvb$, giving
  \[
    \T \vdash \Pvb(\num{\fm{L}}) \leftrightarrow (\Pvb(\num{\Pvb(\num{\fm{L}})}) \rightarrow \Pvb(\num{\fm{A}})).
  \]
  Next, the fact that $\T \vdash \Pvb(\num{\fm{L}}) \rightarrow \Pvb(\num{\Pvb(\num{\fm{L}})})$ (re-representability) and $\T \vdash \Pvb(\num{\fm{A}}) \rightarrow \fm{A}$ (assumption) combines with the previous to produce
  \[
    \T \vdash \Pvb(\num{\fm{L}}) \rightarrow \fm{A}.
  \]
  This, of course, establishes $\T \vdash \fm{L}$ by equivalency which in turn yields $\T \vdash \Pvb(\num{\fm{L}})$ via the representability of $\Pvb$ and thus finally $\T \vdash \fm{A}$ by detachement from the above.
\end{prf}

\begin{theorem}[Second Incompleteness Theorem]
  Let $\T$ be a computably enumerable extension of $\PA$. Then there exists a consistency statement of the form
  \[
    \Con \enskip\coloneqq\enskip \neg \exists \fm{x} \colon \Pvb(\fm{x}) \land \Pvb(\num{\neg} \cat \fm{x})
  \]
  where $\Pvb$ is a strong provability predicate for $\T$. And furthermore, if $\T$ is consistent, then any such statement is unprovable in $\T$.
\end{theorem}

\begin{prf}
  The construction of $\Con$ directly falls out of our previously demonstrated \hyperref[thm-existence]{Existence Theorem}.

  Suppose then for the sake of contradiction that $\T \vdash \Con$. And set $\fm{\top}$ to be some provable sentence, e.g., $\qu{0 \fmeq 0}$. Then by $\Pvb$'s representability condition $\T \vdash \Pvb(\num{\fm{\top}})$ which together with $\T \vdash \Con$ implies $\T \vdash \neg \Pvb(\num{\neg \fm{\top}})$. From this we can get $\T \vdash \Pvb(\num{\neg \fm{\top}}) \rightarrow \neg \fm{\top}$. Hence, by \hyperref[thm-loeb]{Löb's Theorem} $\T \vdash \neg \fm{\top}$; an inconsistency.
\end{prf}

\subsection{A Second Second Incompleteness Theorem (And A Third)}

Despite our innovations engineering a strong provability predicate was still a lengthy undertaking. One of the rewards was that we got to visit Löb's Theorem on our way towards the Second Incompleteness Theorem. More importantly though, there's an air of authority surrounding the HBL-derivability conditions. They offer a stamp of approval that our consistency statement is \emph{proper} in some sense. However, there's a result due to Robert G. Jeroslow \cite{jeroslow} which trades away the distributivity condition in return for a slightly stronger variant of the re-representability condition and which still manages to achieve the Second Incompleteness Theorem. And this result is happily cited in the literature from time to time. It would appear then that formulations of the Second Incompleteness Theorem giving less assurances about their consistency statement are still legitimate. Just how far can we conceivably lower this bar? Let's test those limits.

The key is to realize that demanding the distributivity and re-representability condition for arbitrary sentences is to use a sledgehammer to crack a nut. What we require is merely to make our predicate see the truth of certain arguments involving $\fm{G}$. This is something we can accomplish with far less surgery.

\begin{theorem}[Little Second Incompleteness Theorem---Gödel's Version]
  Let $\T$ be a consistent and computably enumerable theory which correctly represents every computable function. Then there exists an unprovable consistency statement of the form
  \[
    \Con \enskip\coloneqq\enskip \neg \exists \fm{x} \colon \Pvb(\fm{x}) \land \Pvb(\num{\neg} \cat \fm{x})
  \]
  where $\Pvb$ is a weak provability predicate for $\T$.
\end{theorem}

\begin{prf}
  Using the \hyperref[lm-diag]{Diagonal Lemma} on $\neg\PvbG$ we can obtain an unprovable Gödel sentence $\fm{G}$ and then put
  \[
    \Pvb(\fm{x}) \enskip\coloneqq\enskip \Pvb_\fm{G}(\fm{x}) \lor \neg\fm{G}.
  \]

  First, let's check that $\Pvb$ is indeed a weak provability predicate. We know from discussing the \hyperref[thm-first]{First Incompleteness Theorem} that $\neg\fm{G}$ is false. Therefore, $\Pvb(\num{\fm{A}})$ is true if and only if $\Pvb_\fm{G}(\num{\fm{A}})$ is true, allowing $\Pvb$ to effortlessly inherit $\Pvb_\fm{G}$'s correctness property. Representability is similarly trivial.

  And as for the unprovability of $\Con$, just expanding definitions and simple logic reveals $\Con \leftrightarrow (\Con_\fm{G} \land \fm{G})$. Thus, we have $\Con \rightarrow \fm{G}$, making $\Con$ unprovable because $\fm{G}$ is.
\end{prf}

The predicate $\Pvb$ used in the above proof probably appeared to some readers like a rabbit out of a hat. So let me try to motivate it a little further. Using the equivalency $\fm{G} \leftrightarrow \neg\Pvb_\fm{G}(\num{\fm{G}})$ we can rewrite $\Pvb(\fm{x})$ as $\Pvb_\fm{G}(\fm{x}) \lor \Pvb_\fm{G}(\num{\fm{G}})$. Without disrupting the argument this can then be complicated further into
\[
  \Pvb(\fm{x}) \enskip\colonapprox\enskip (\neg\, \fm{x} \fmeq \num{\neg\fm{G}} \land \Pvb_\fm{G}(\fm{x})) \lor (\fm{x} \fmeq \num{\neg\fm{G}} \land \Pvb_\fm{G}(\num{\fm{G}})).
\]
Now let me antropomorphize $\Pvb$ once again. Whenever it encounters an ordinary sentence $\Pvb$ simply turns around and queries $\Pvb_\fm{G}$. However, when presented with $\neg\fm{G}$ our predicate reasons: ``Oh, you want to know whether I believe $\neg\fm{G}$? Well, that's the same as asking whether I believe $\Pvb_\fm{G}(\num{\fm{G}})$. Certainly I'd believe that if $\Pvb_\fm{G}$ assured me of $\fm{G}$.'' The intuition is exactly the same as it is behind the Existence Theorem. It has only been narrowed down to a special case and stripped of all redundancies.

Notice also that if one actually managed to complete the herculean task of demonstrating the derivability conditions for $\Pvb_\fm{G}$, then $\Pvb$ would be able to inherit those just as well. And ultimately, after confirming $\Con_\fm{G} \rightarrow \fm{G}$, we'd come full circle to reveal $\Con \leftrightarrow \Con_\fm{G}$. You could say that all we've done is take a drastic shortcut.

Finally, there's one more obvious idea we need to try. Actually, you try it.

\begin{theorem}[\phantom{Little Second Incompleteness Theorem---Rosser's Version}]
  Let $\T$ be a consistent and computably enumerable extension of $\PA$. Then there exists an \emph{undecidable} consistency statement of the form
  \[
    \Con \enskip\coloneqq\enskip \neg \exists \fm{x} \colon \Pvb(\fm{x}) \land \Pvb(\num{\neg} \cat \fm{x})
  \]
  where $\Pvb$ is a weak provability predicate for $\T$.
\end{theorem}

\begin{prf}
  Exercise.
\end{prf}

\section{A Hierarchy Theorem}

Next up, I want to broaden our horizons a little by looking at some more novel connections to complexity theory. Truth be told, this section was the original motivation for writing all of this. I just couldn't bring myself to beeline straight to this point without filling in the bigger picture. In the process of doing so one thing led to another till at this point I consider this section to possibly be the least interesting. At the same time, I didn't want to leave it on the cutting room floor.

The thing is, similarities between the First Incompleteness Theorem and the Halting Problem have long been noted. But looking at our simplified proof of the Incompleteness Theorem they stand out more than ever. The Halting problem is proved undecidable by running a certain program on its own source code. And the First Incompleteness Theorem follows from running a very similar program on its own representation.

This might bring to mind then that the Halting Problem was famously modified by Juris Hartmanis and Richard E. Stearns in 1965 in order to establish the Time Hierarchy Theorem \cite{hartmanis-stearns}. A question thus arises: can the same modification be performed on the First Incompleteness Theorem? This is something Gödel himself already undertook for his so called Speedup Theorem \cite{speedup}. But by generalizing to sets of sentences and adding a padding trick we can take this idea even further and establish an analogue of the Nondeterministic Time Hierarchy Theorem, originating from Stephen A. Cook \cite{cook}. Here's how.

To start with we're going to have to be more restrictive as to what constitutes a proof. Mere computable enumerability doesn't cut it anymore.

\begin{definition}[Canonical Proof Verifier]
  A proof verifier $\pc$ for some theory $\T$ is considered a \emph{canonical proof verifier} if for any $p \in \mathbb{N}$ such that $\pc(p, \fm{A}) = \cmark$ for some sentence $\fm{A}$ the following holds:
  \begin{itemize}
    \item $p$ is a list containing only theorems of $\T$ or valid sub-proofs.
    \item If $\fm{B}$ follows from the items in $p$ by the usual rules of inference, then $\pc(p \Cat [\fm{B}], \fm{B}) = \cmark$.
  \end{itemize}
\end{definition}
I'm neglecting to fully spell out the details of the above definition because, as we'll see, all we're going to need is the ability to turn a proof of $\fm{A}$ and a proof of $\fm{B}$ into a proof of $\fm{A} \land \fm{B}$ and vice versa. I hope you'll grant that the above definition covers that.

We'll also need to define the analogue of a computational problem.

\begin{definition}[String Length]
  $\lvert \cdot \rvert$ denotes the computable function which determines the length of a number. (The number of digits required to write it.)
\end{definition}

\begin{definition}[Proof Problem]
  Fix a theory $\T$ and a proof verifier $\pc$. Then any set of sentences $\A \subseteq \T$ will be considered a \emph{proof problem}. Furthermore, the \textit{(proof) complexity} $f$ of $\A$ is defined to be
  \[
    f(n) \coloneqq \max_{\substack{\enskip\fm{A} \in \A \colon\\ \lvert \fm{A} \rvert \leq n}} \; \min_{\substack{p \in \mathbb{N} \colon\\ p \text{ $\pc$-decides } \fm{A}}} \lvert p \rvert
  \]
\end{definition}
In plain English, the complexity of a proof problem the minimum number of symbols required to decide every sentence up to length $n$.

Onto the result.

\begin{theorem}[Proof Hierarchy Theorem]
  Let $\T$ be a consistent and computably enumerable theory which represents every computable function. Let $g$ be computable, monotone, and $g(n) = \Omega(n)$. And let $f(n+1) = o(g(n))$. Then for any canonical proof verifier $\pc$ there exists a proof problem with complexity $O(g(n))$ but not $O(f(n))$.
\end{theorem}

\begin{prf}
  Consider the following algorithm.
  \begin{algo}
    \Function{$Y(\rep{f}, x)$}{}
    \State $m \coloneqq \bigl\lvert \neg\, f(\numrep{f}, \num{x}) \fmeq \num{\cmark}\bigr\rvert$
    \For{$p\in\mathbb{N} : \lvert p \rvert \leq g(m)$}
    \If{$p$ $\pc$-proves $\neg\, f(\numrep{f}, \num{x}) \fmeq \num{\cmark}$}
    \State\Return $\,\cmark$
    \EndIf
    \If{$p$ $\pc$-proves $\phantom{\neg\,} f(\numrep{f}, \num{x}) \fmeq \num{\cmark}$}
    \State\Return $\xmark$
    \EndIf
    \EndFor
    \State\Return $\xmark$
    \EndFunction
  \end{algo}
  This gives rise to the proof problem
  \[
    \B \enskip\coloneqq\enskip \Bigl\{\neg\, Y(\numrep{Y}, \num{x}) \fmeq \num{\cmark} : x \in \mathbb{N} \Bigr\}
  \]
  where the $x$ just serves to turn our single undecidable sentence from the First Incompleteness Theorem into an infinitely large family.

  To start with, $\B$ can't have complexity $g(m)$ by reasoning exactly as in the First Incompleteness Theorem. Any proofs this short would be found by $Y$ and swiftly lead to an inconsistency. That means however that \emph{eventually} $Y$ will arrive at its last line and output $\xmark$. By the \hyperref[lm-run]{Runtime Lemma} that makes every sentence in $\B$ provable and gives this problem a well defined proof complexity $h$ such that $g(m) < h(m)$.

  Now let $g^{-1}(j) \coloneqq \argmin_i j \leq g(i)$ and define
  \[
    \A \enskip\coloneqq\enskip \Bigl\{ \fm{B} \land \neg\, \fm{S} \dots \fm{S}\fm{0} \fmeq \fm{0} : \fm{B} \in \B,\; n = g^{-1}(h(m)) \Bigr\}
  \]
  where $m$ is the length of the original sentence $\fm{B} \in \B$ and $n$ is the length of the new sentence $\fm{A} \in \A$. In other words, we add $\fm{S}$'s until we hit $n = g^{-1}(h(m))$. The aim here is to dilute the sentences in $\B$ with padding until we reach complexity $O(g(n))$ and no lower. Also, recall that by virtue of $\T$ representing at least \emph{some} function we have the axioms of successorship available.

  First, let's verify that $\A$ has proof complexity $O(g(n))$. To that end note that $i \leq g(g^{-1}(i))$ for any $i \in \mathbb{N}$. So we can say that it takes $h(m) \leq g(g^{-1}(h(m))) = g(n)$ symbols to prove any $\fm{B} \in \B$. And it takes $O(n)$ symbols to prove the padding. Since $g$ is monotone and $g(n) = \Omega(n)$ this gives us proofs of length $O(g(n))$ overall.

  And second, we need to rule out $O(f(n))$ as a proof complexity for $\A$. This time, opposite to the previous case, note that $g(g^{-1}(i) - 1) < i$ for any $i \in \mathbb{N}$. Hence,
  \[
    O(f(n)) = o(g(n-1)) = o(g(g^{-1}(h(m))-1)) = o(h(m)).
  \]
  This leads us to conclude that if $\A$ and thereby $\B$ had proofs of length $O(f(n))$, then it would have proofs of length $o(h(m))$. But that contradicts the assumption that $h(m)$ denoted the length of the shortest proofs for $\B$.
\end{prf}

As already mentioned, this theorem is quite similar to the Nondeterministic Time Hierarchy. Here's a reminder.

\begin{theorem}[Nondeterministic Time Hierarchy Theorem]
  Let $g$ be time constructible (one can compute $g(n)$ in $O(g(n))$ steps). And let $f(n+1) = o(g(n))$. Then $\mathsf{NTIME}(f(n)) \subsetneq \mathsf{NTIME}(g(n))$.
\end{theorem}

\begin{prf}
  Omitted.
\end{prf}

There are a couple of lines along which I'd like to compare these theorems.

\begin{itemize}
  \item Fascinatingly, even though the proof of the Proof Hierarchy Theorem runs along very different lines than the proof of the Nondeterministic Time Hierarchy Theorem (at least the one I'm familiar with by \v{Z}ák Stanislav \cite{stanislav}), the exact same $+1$ shows up in the formulation.
  \item Whereas the Nondeterministic Time Hierarchy Theorem requires $g$ to be time constructible, the Proof Hierarchy Theorem is already satisfied with mere computability.
  \item The Proof Hierarchy Theorem doesn't make any demands regarding the computational complexity of checking proofs. In particular checking whether a given sentence is an axiom might be arbitrarily difficult.
  \item Fundamentally, the nondeterminstic time hierarchy is a hierarchy \emph{among} verifiers, while the proof hierarchy is a hierarchy among the inputs for a \emph{fixed} verifier.
  \item The main sticky point that prevents this proof from being turned into a simpler demonstration of the Nondeterministic Time Hierarchy Theorem is that, as Manuel Blum demonstrated \cite{blum}, computational problems might not have a \emph{minimum} complexity $h$.
\end{itemize}

\section{Oracles}

Last but not least, I want to take a brief look at a topic near and dear to computer scientists of the present: oracles. This section was born from the following line of thought.

Take Peano Arithmetic and add a set of uncomputable axioms to it. The simplest example would be an axiomatization of the Halting Problem by means of a predicate $\mathsf{H}$ so that either $\mathsf{H}(\num{x})$ or $\neg\mathsf{H}(\num{x})$ is an axiom depending on whether or not $x$ describes a halting computation. This theory, call it $\PA+\mathsf{H}$, is no longer computably enumerable so the First Incompleteness Theorem won't apply anymore. The step that breaks in the proof is that the algorithm $X$ we constructed now needs to run an uncomputable proof verifier as a subroutine. Sure, we could make $X$ ``computable'' if we gave it access to an oracle for the Halting Problem. But then we still face the challenge of representing $X$. We only know how to represent computable functions, not functions that call on a Halting Problem oracle. To represent such functions it seems that we'd also need a way to represent the Halting Problem. Except, that is precisely what $\PA+\mathsf{H}$ conveniently provides for us!

To turn this idea into a proof we're going to need a lemma. This one:

\begin{lemma}[Oracle Representation Lemma]
  Let $\T$ be an extension of $\Q$ which also represents $g$. Then $\T$ represents any function $f$ that can be computed using oracle access to $g$.
\end{lemma}

\begin{prf}
  First we set up the following pair of algorithms.
  \begin{algo}
    \Function{$\fcheck(\bm{x}, p)$}{}
    \State \textbf{simulate} $f$ running on $\bm{x}$ using $p$ in place of the oracle for $g$
    \If $p$ isn't of the form $[[q_1, r_1], \dots, [q_k, r_k]]$ \textbf{or} some $q_i$ isn't a query $f$ would've made
    \State \Return $\xmark$
    \EndIf
    \State \Return $\cmark$
    \EndFunction
  \end{algo}
  \begin{algo}
    \Function{$\fsim(\bm{x}, p)$}{}
    \State \textbf{simulate} $f$ running on $\bm{x}$ using $p$ in place of the oracle for $g$
    \State \Return $f(\bm{x})$
    \EndFunction
  \end{algo}
  It'll also be helpful to use a shorthand:
  \[
    \mathsf{\Chk(\bm{\mathsf{a}}, p) \enskip\coloneqq\enskip \fcheck(\bm{\mathsf{a}}, p) \fmeq \num{\cmark} \land \forall i < \lVert p \rVert \colon \uq{g}(p[i][0]) \fmeq p[i][1]}.
  \]
  Then we can give this representation of $f$:
  \[
    \rep{f}(\bm{\mathsf{a}}, \mathsf{b}) \enskip\coloneqq\enskip \mathsf{\exists p \colon \fsim(\bm{\mathsf{a}}, p) \fmeq b \land \Chk(\bm{\mathsf{a}}, p) \land \forall q < p \colon \neg \Chk(\bm{\mathsf{a}}, q)}.
  \]

  To verify that $\rep{f}$ works as advertised suppose that $f(\bm{x}) = y$. Then there exists a shortest corresponding list $p_y$ of oracle inputs and outputs for $g$. This allows us to confirm the following two required statements.
  \begin{description}
    \item[$\T \vdash \rep{f}(\num{\bm{x}}, \num{y})$]
          This easily follows from
          \[
            \mathsf{\fsim(\num{\bm{x}}, \num{p_y}) \fmeq \num{y} \land \Chk(\num{\bm{x}}, \num{p_y}) \land \forall q < \num{p_y} \colon \neg \Chk(\num{\bm{x}}, q)}.
          \]
          The bounded quantifiers can be converted into a finite number of cases using the \hyperref[lm-bounded]{Bounded Quantification Lemma} (or the \hyperref[lm-noneg]{non-existence of negative objects} if necessary). And then it's just a matter of proving via the \hyperref[lm-run]{Runtime Lemma} that a bunch of representable functions when run on the given inputs produce the given outputs.
    \item[$\T \vdash \qu{\exists! y \colon \rep{f}(\num{\bm{x}}, y)}$]
          Existence has already been handled. For uniqueness let $\fmc{y}$ be arbitrary and assume $\rep{f}(\num{\bm{x}}, \fmc{y})$. Then we can obtain a $\fmc{p}$ such that
          \[
            \qu{\uq{\fsim}(\num{\bm{x}}, \fmc{p}) \fmeq \fmc{y} \land \Chk(\num{\bm{x}}, \fmc{p}) \land \forall q < \fmc{p} \colon \neg \Chk(\num{\bm{x}}, q)}.
          \]
          Of course, we also still have
          \[
            \qu{\uq{\fsim}(\num{\bm{x}}, \num{p_y}) \fmeq \num{y} \land \Chk(\num{\bm{x}}, \num{p_y}) \land \forall q < \num{p_y} \colon \neg \Chk(\num{\bm{x}}, q)}
          \]
          which allows us to easily rule out $\fmc{p} < \num{p_y}$ and $\num{p_y} < \fmc{p}$. By the \hyperref[lm-total-num]{totality of ordering w.r.t. numerals} this leaves $\fmc{p} \fmeq \num{p_y}$ and hence
          \[
            \qu{\fmc{y} \fmeq \fsim(\num{\bm{x}}, \fmc{p}) \fmeq \fsim(\num{\bm{x}}, \num{p_y}) \fmeq \num{y}}.\qedhere
          \]
  \end{description}
\end{prf}

We're now ready to answer the question regarding the completeness of $\PA + \mathsf{H}$ we raised earlier. After all, this theory can represent the Halting Problem through the formula $(\fm{H}(\fm{x}) \land \fm{y} \fmeq \num{\cmark}) \lor (\neg \fm{H}(\fm{x}) \land \fm{y} \fmeq \num{\xmark})$. So using the Oracle Representation Lemma it can represent the algorithm $X$ used in the proof of the First Incompleteness Theorem and the argument goes through.

Unquestionably, there's a general theorem to be proved here. Actually, many of them are. For the assumption of computable enumerability can now be relaxed throughout our entire discussion. But relaxed to what? What exactly characterizes theories such as $\PA + \mathsf{H}$? One option might be ``a theory which represents its own proof verifier''. But that assumption is already doing half the work towards establishing our generalization. There has to be something more natural.

In any case, I need to rest the pen eventually. And this seems like a good place to take a break.
\vspace{-0.2cm}

\epigraph{The final answer I badly wanted to blurt out was: ``No, no, no---Gödel \emph{did} invent Lisp!''}{\textit{Douglas Hofstadter}}

\vfill\eject

\end{document}